\documentclass[fleqn,usenatbib]{mnras}

\usepackage{newtxtext,newtxmath}

\usepackage[T1]{fontenc}

\DeclareRobustCommand{\VAN}[3]{#2}
\let\VANthebibliography\thebibliography
\def\thebibliography{\DeclareRobustCommand{\VAN}[3]{##3}\VANthebibliography}

\usepackage{graphicx}	
\usepackage{amsmath}	

\usepackage{enumitem}
\usepackage{booktabs}
\usepackage{xspace}
\usepackage{textcomp}
\usepackage{xcolor}	
\usepackage[euler]{textgreek}
\usepackage{dirtytalk}
\usepackage{tabularx}
\usepackage{caption}
\usepackage{subcaption}
\usepackage{natbib,twoopt}
\usepackage{hyperref}

\graphicspath{{./figures}}
\graphicspath{{./appendix_figures}}

\newcommand{\Msun}{{\rm M}_{\odot}}

\newcommand{\Mstar}{M_*}
\newcommand{\Halpha}{H$\alpha$}
\newcommand{\msafit}{{\tt msafit}}
\newcommand{\pysersic}{{\tt pysersic}}
\newcommand{\Sersic}{S{\' e}rsic}
\newcommand{\msavre}{$v(r_{\rm e})$}
\newcommand{\msasigma}{$\sigma_0$}
\newcommand{\TNGvre}{$v_{\rm max}(r\leq r_{\rm e})$}
\newcommand{\TNGsigma}{$\sigma_e$}
\newcommand{\kms}{\rm km\,s^{-1}}
\newcommand{\orcid}[1]{\href{https://orcid.org/#1}{\includegraphics[scale=0.25]{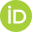}}}

\title[Testing the inference of kinematics from mock NIRSpec/MSA observations]{Testing the inference of kinematics from mock JWST NIRSpec/MSA observations of TNG50 galaxies at $z\sim2-6$}

\author[R. Anirudh et al.]{
Ravishankar Anirudh$^{1,2}$\orcid{0000-0002-3071-3365}\thanks{E-mail: ravishankar@mpia.de\newline Fellow of the International Max Planck Research School for Astronomy and Cosmic Physics at the University of Heidelberg (IMPRS-HD)},
Anna de Graaff$^{3,1}$\orcid{0000-0002-2380-9801}\thanks{Clay Fellow},
Florian Lacroix$^{1}$,
Sedona H. Price$^{4}$\orcid{0000-0002-0108-4176},
\newauthor\
and Annalisa Pillepich$^{1}$\orcid{0000-0003-1065-9274}
\\
$^{1}$Max-Planck-Institut f{\"u}r Astronomie, K{\"o}nigstuhl 17, D-69117 Heidelberg, Germany\\
$^{2}$Fakult{\"a}t f{\"u}r Physik und Astronomie, Universit{\"a}t Heidelberg, Im Neuenheimer Feld 226, 69120 Heidelberg, Germany\\
$^{3}$Center for Astrophysics $|$ Harvard \& Smithsonian, 60 Garden St., Cambridge MA 02138 USA \\
$^{4}$Space Telescope Science Institute, 3700 San Martin Drive, Baltimore, MD 21218, USA
}

\date{Accepted XXX. Received YYY; in original form ZZZ}

\pubyear{\the\year{}}

\begin{document}
\label{firstpage}
\pagerange{\pageref{firstpage}--\pageref{lastpage}}
\maketitle              

\begin{abstract}
We use the TNG50 galaxy formation simulation to generate mock JWST NIRCam and NIRSpec microshutter array (MSA) observations of \Halpha-emitting gas in $M_*=10^8-10^{11.5}\,\Msun$ star-forming galaxies at $z=2-6$. We measure morphological properties from the mock imaging through \Sersic\ profile fitting, and gas rotational velocities ($v$) and velocity dispersions ($\sigma$) by fitting the mock spectra as thin, rotating discs. To test the efficacy of such simple parametric models in describing complex ionised gas kinematics, we compare the best-fit quantities to intrinsic simulation measurements. At $z=3$, we find that $v$ and $\sigma$ for aligned and resolved sources generally agree well with intrinsic measurements, within a factor of $\sim$2 and $\sim$1.5, respectively. The recovery of kinematics is robust for smooth, disc-like systems, but $v$ and $\sigma$ can be over- or underestimated by more than a factor of 2, respectively, for intrinsically elongated systems. The scatter in the recovery accuracy is larger at higher redshift, as TNG50 galaxies at $z>3$ deviate more strongly from the thin rotating disc assumption. Despite uncertain measurements for individual galaxies, we find that key population trends, such as the weak redshift evolution of $\sigma$ and $v/\sigma$ as well as the dependence of $\sigma$ on the global star formation rate, are broadly recovered by our kinematic modelling. Our work provides the end-to-end framework needed to compare NIRSpec MSA observations to cosmological simulations and to quantify observational biases in measuring ionised gas kinematics, highlighting the need for the development of dedicated models for high-redshift galaxies.
\end{abstract}




\section{Introduction}
    \label{sec:intro}
    
    Star-forming galaxies evolve from gas-rich structures in the early Universe to the stellar discs commonly seen today. A key open question is when and how galaxies settle into these disc-like structures. Observationally, this is probed by measuring the dynamics of gaseous and stellar emission from galaxies across cosmic time. In contrast, hydrodynamical simulations of galaxy formation typically provide 6-dimensional spatial and kinematic information of gas and stars \citep{Vogelsberger_2020,Crain_2023}, yet often without explicitly solving for radiation from various galaxy components in a full cosmological context \citep[but see e.g.][]{Katz_2022,Bhagwat_2024,Kannan_2025}. Bridging the gap between numerical outcomes and observables thus requires the post-processing of simulation data, e.g., to produce emission maps akin to observations.
    
    Although spatially-resolved stellar dynamical measurements are generally limited to galaxies at $z\lesssim2$ \citep{vandeSande_2013,vanHoudt_2021,Slob_2025,Rhoades_2025} \citep[yet see also][]{Newman_2018}, bright emission lines from warm ionised gas and cold gas reveal the motions of gas within galaxies in the early Universe beyond cosmic noon up to $z\sim7$ \citep{Rowland_2024,Ivey_2025}, with tentative signatures of rotation even at $z\sim14$ \citep{Scholtz_2025}. The key quantities measured with these spectroscopic observations are rotational velocities ($v$) and velocity dispersions ($\sigma$) of the gas, typically within the aperture of the size of the galaxy. Together, these two quantities describe whether galaxies are supported primarily by rotation or pressure, classifying them as dynamically cold or hot systems based on $v/\sigma$.

    Simulations of galaxy formation such as IllustrisTNG \citep{Nelson_2018,Pillepich_2019} predict that star-forming galaxies of $M_*\sim10^{10}\,\Msun$ become more pressure-supported toward higher redshifts, going from $v/\sigma\sim6$ at $z=2$ to $v/\sigma\leq3$ for $z\geq4$ (on average). However, models differ in their estimates of the rotational support of high-redshift galaxies. For instance, \citet{Kohandel_2024} used the SERRA simulations combined with post-processed [CII] and \Halpha\ emission maps to predict the existence of massive ($M_*>10^{10}\,\Msun$) dynamically cold discs with $v/\sigma>5$ and $\sigma\lesssim100$ km $\mathrm{s^{-1}}$ even up to $z\sim7$. These studies highlight the connection of gas velocity dispersions to the stability of gas discs against different sources of energy, i.e., feedback and gravitational instabilities \citep{Toomre_1964,Krumholz_2018}. Theoretical studies have shown that stellar feedback, mass transport, and cosmological inflows all impact the evolution of gas turbulence \citep{Genel_2012,Orr_2020}. To disentangle these effects and robustly differentiate between galaxy formation models, it is crucial to measure gas velocities and dispersions for galaxies that span not only a broad range of cosmic time but also exhibit a wide variety of galaxy properties.
    
    Large ground-based spectroscopic surveys of the ionised gas kinematics, such as SINS, KMOS3D, and MOSDEF, have together measured $v$ and $\sigma$ for over a thousand galaxies at $z\sim1-3$ \citep{Foerster_Schreiber_2009,Kriek_2015,Wisnioski_2019,Price_2020}. A key finding from these efforts, particularly with the \Halpha\ emission line, is the systematic increase of (ionised) gas turbulence in galaxies from $z\sim0.5-3$ \citep{Uebler_2019,Wisnioski_2019}. At higher redshifts, sub-millimetre studies in the last decade with large telescopes such as the Atacama Large Millimeter Array (ALMA) have revealed the dynamics of several tens of massive ($M_*\geq10^{10}\,\Msun$) star-forming galaxies \citep{Neeleman_2020,Fraternali_2021,Lelli_2021} using gas tracers such as CO, [CI], and [CII]. Dynamically cold systems have been observed up to $z=7$ \citep{Smit_2018,Rizzo_2021,Rowland_2024,Lee_2025b}, suggesting significantly earlier formation times for galaxies at these masses than expected. However, these sub-millimetre studies have primarily focused on the most massive systems with high star formation rates \citep{Neeleman_2020,Fraternali_2021}, due to the limited sensitivity of ALMA. Only few strongly-lensed sources with lower stellar masses at $z>4$ have been studied to date \citep{Pope_2017,Pope_2023}, but the selection function of sources that benefit from strong lensing is complex. To understand the dynamical evolution of the galaxy population, it is therefore crucial to sample more typical galaxies ($M_*<10^{10}\,\Msun$) at $z\geq2$.
    
    With the excellent sensitivity and resolution of JWST, it is now viable to measure spatially resolved ionised gas kinematics from \Halpha\ and [O\textsc{iii}] emission in individual galaxies up to $z\sim7.5$ \citep{deGraaff_2024,Parlanti_2024,Arribas_2024,Ivey_2025}, with statistical estimates of $v/\sigma$ for low-mass galaxies extending to $z\geq4$ \citep{Danhaive_2025a}. These studies probe, for the first time, the galaxy population at stellar masses of $M_*=10^7-10^9\Msun$, which are the progenitors of the more massive galaxies observed at cosmic noon \citep{Genzel_2020,Wang_2025}.
    
    Much of this success can be attributed to spectroscopy with the JWST NIRSpec spectrograph \citep{Gardner_2023,Jakobsen_2022}. Although Integral Field Unit (IFU) observations provide a more detailed view of galaxy kinematics, the capability of the NIRSpec microshutter array (MSA) to obtain spectra for hundreds of sources efficiently enables a statistical analysis of rotational support across cosmic epochs. However, spatial information is inherently limited for the MSA, thereby trading it for sample size instead, compared to the IFU. Furthermore, the complexity of the NIRSpec instrumental Line Spread Function (LSF) and the Point Spread Function (PSF) requires a dedicated forward modelling approach to infer kinematic properties from these spectra \citep{Jakobsen_2022}.
    
    Forward modelling tools for slit-based spectrographs generate mock 2D spectra based on a parametrisation of the morphology of the emission line and the underlying velocity field, which are then fit to observed spectra \citep{Price_2016,Straatman_2022,deGraaff_2024}. Models typically assume galaxies to be axisymmetric, having fixed disc thickness (or even none), and be described by \Sersic\ intensity profiles \citep{Sersic_1968}, arctangent velocity profiles \citep{Courteau_1997} and a constant velocity dispersion profile. However, these model assumptions may be invalid in some cases - for instance, spheroidal galaxies are common in the early Universe, especially at lower stellar masses \citep{Pillepich_2019}, thereby rendering thin disc geometries inaccurate. This raises important questions on whether relatively simple kinematic models could describe the great complexity of realistic galaxies, how informative they may be, and when they might fail.
    
    Cosmological simulations are well-suited to ascertain these observational biases, as they provide a large sample of galaxies with realistic physical properties, thereby enabling testing of inferences from kinematic modelling. In this context, \citet{Uebler_2019} inferred galaxy kinematics from mock IFU maps generated using TNG50 and compared them with ground-based observations \citep{Genzel_2020}, yielding minor differences in \Halpha-inferred and intrinsic dark matter fractions for seven $z\sim2$ galaxies. More recently, \citet{Phillips_2025} created synthetic NIRSpec IFU \Halpha\ observations of two massive $z>9$ galaxies from the SERRA simulations, finding biases in inferred velocity dispersions beyond a factor of $\sim2-3$ compared to the intrinsic values. These biases stem from overestimations of $\sigma$ due to the presence of inflows and outflows of ionised gas. Tracers of different gas phases, such as \Halpha\ (warm ionised) and [CII] (cold neutral or molecular), can therefore provide varying estimates of the velocity dispersion \citep{Kohandel_2024}. Despite these known shortcomings, the lack of statistical comparisons between observations and forward-modelled simulated galaxies, particularly for the NIRSpec/MSA, impedes a clear understanding of possible lapses in galaxy formation physics. Consequently, they emphasise the need to validate whether inferred kinematics accurately represent the underlying galaxy population.

    To this end, we use TNG50, which provides large samples of simulated galaxies with high resolution, thereby enabling statistical comparisons of JWST NIRSpec observations and forward-modelled simulations. In this work, we construct the first realistic mock NIRSpec/MSA observations for a sample of $>4000$ galaxies of $M_*>10^8\,\Msun$ from the TNG50 simulation across $2\leq z\leq6$. We use these mock observations to test the inference of galaxy kinematics by modelling them as thin, rotating discs. A key outcome of this work is to assess if observational inferences can accurately recover meaningful estimates of the kinematics of high-redshift galaxies, and where they may be biased or approximations fail.

    We describe the simulations and their intrinsic measurements of $v$ and $\sigma$ in Sec.~\ref{subsec:sims} and~\ref{subsec:TNG50_intrinsic}. The methodology for making \Halpha\ emission maps and 2D spectra are presented in Sec.~\ref{subsec:mock}. The procedure for inference of morphological and kinematic properties from these mocks is demonstrated in Sec.~\ref{subsec:inference}. In Sec.~\ref{sec:inference_results}, we present the results on best-fit galaxy kinematics, their possible biases in relation to intrinsic quantities, and their evolution with cosmic time. We comment on velocity dispersion scaling relations for best-fit versus intrinsic measurements in Sec.~\ref{sec:sigma_scaling}. We discuss the implications of our findings in Sec.~\ref{sec:discussion} and summarise our work in Sec.~\ref{sec:conclusions}.

\section{Methodology}

    Our procedure for generating synthetic observations from simulated galaxies and estimating their mock-observed morphological and kinematic properties is as follows. We use the TNG50 cosmological galaxy formation simulation to extract star-forming galaxies of interest (Sec.~\ref{subsec:sims}). We then create mock JWST NIRCam photometric images (Sec.~\ref{sssec:nircam}) and mock JWST NIRSpec/MSA spectra (Sec.~\ref{sssec:nirspec}). We extract morphological properties of the \Halpha-emitting gas from the mock photometric images using \pysersic\ (Sec.~\ref{sssec:sersic}), assuming the \Halpha\ morphology follows the continuum emission. We then use these morphological properties as priors to fit for the kinematics of each galaxy assuming a thin, rotating disc model (Sec.~\ref{sssec:kin_model}). In this work, we focus primarily on the testing of kinematic inference, rather than commenting on the validity of the simulation models themselves.
    
    \subsection{The TNG50 simulation}
        \label{subsec:sims}
    
    We analyse galaxies from the TNG50-1 simulation \citep[hereafter, TNG50][]{Pillepich_2019,Nelson_2019a,Nelson_2019b}, the highest resolution volume of the IllustrisTNG project \citep{Marinacci_2018,Naiman_2018,Nelson_2018,Springel_2018,Pillepich_2018b}. All IllustrisTNG simulations are run with the moving-mesh code {\sc AREPO} \citep{Springel_2018} assuming a $\Lambda$CDM cosmology with cosmological parameters from \citet{Planck_2016}. They are based on the IllustrisTNG galaxy formation model \citep{Weinberger_2017,Pillepich_2018a}, which include prescriptions for star formation, stellar evolution, the growth of supermassive black holes (SMBHs), feedback from supernovae and SMBHs.
    
    The TNG50 volume of $\sim$50$^3$ comoving Mpc$^3$, combined with the baryonic mass resolution $m_{\rm baryon}$ of $8.4\times10^4\,\Msun$, allows for the analysis of thousands of galaxies at high spatial resolution, with typical (star-forming) gas cell sizes of $<$100 pc at $z=2-4$. These gas cell sizes are a factor of 5 or more smaller than the NIRCam PSF FWHM at the \Halpha\ rest-frame wavelength at these redshifts \citep[see for e.g. Fig.~1 of][]{Pillepich_2019}. In addition, the dark matter particle resolution $m_{\rm DM}$ of $4.5\times10^5\,\Msun$ and Plummer-equivalent softening lengths of $0.29$ kpc for stars and dark matter at $z=0$ enables the robust tracing of the gravitational potential across a wide range of galaxy stellar masses.

    Galaxies are identified using the {\sc SUBFIND} algorithm \citep{Springel_2001}. Gas cells in TNG50 are tagged as star-forming when their density exceeds a threshold of $n_{\rm H}=0.13\,\mathrm{cm^{-3}}$. These gas cells are placed on an ``effective equation of state" \citep[as described in][]{Springel&Hernquist_2003,Pillepich_2018a}, i.e., their temperatures are fixed analytically via their densities. In this work, we select galaxies with a star formation rate (SFR) $>0$ \citep[see][for a discussion on vanishing SFR values from IllustrisTNG]{Donnari_2019} and a stellar mass $\Mstar\geq10^8\,\Msun$, both defined within twice the 3D stellar half-mass radius ($r_{\rm 1/2,*}$). Given the relatively small box size of TNG50 (in comparison to TNG100 or TNG300), galaxies more massive than $10^{10}\,\Msun$ are rare at $z>6$ in this simulation. Consequently, we focus on four snapshots at $z=2,3,4,6$ which correspond to epochs typically probed by JWST.
        
    \subsection{Intrinsic measurements of TNG50 galaxy kinematics}
        \label{subsec:TNG50_intrinsic}

    A central goal of this work is to test the efficacy of our kinematic modelling of mock spectra (described in Sec.~\ref{subsec:inference}) at inferring the kinematics of simulated galaxies and compare them to intrinsic measures of $v$ and $\sigma$. Measuring the ``intrinsic'' rotation and dispersion for these galaxies, however, strongly depend on the choice of binning, aperture size, weighting, and projection \citep[see Appendix A of][]{Pillepich_2019}. We closely follow the approach of \citet{Pillepich_2019} and make new measurements of the kinematic properties of galaxies in our sample, thereby extending to lower stellar masses than in their work. We describe these steps in detail, including modifications, below.

    \begin{figure}
        \centering
        \includegraphics[width=\columnwidth]{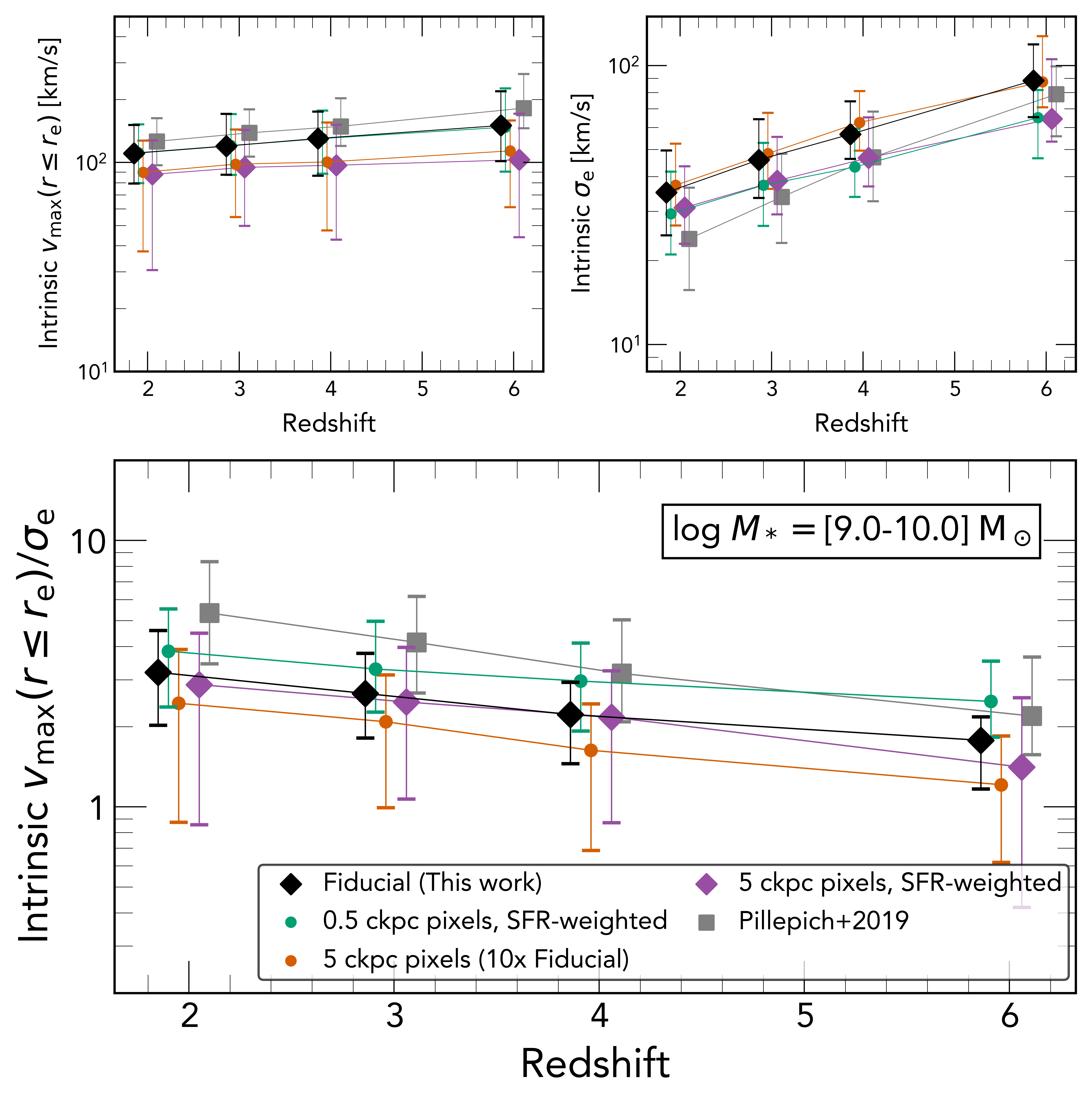}
        \caption{Intrinsic $v$, $\sigma$ and $v/\sigma$ vs redshift for TNG50 galaxies with stellar masses of $10^8-10^{11.5}\,\Msun$. The markers and the errorbars correspond to the median and the 16th and 84th percentiles, respectively. Different colours correspond to various measurement choices. Measurements from \citet[][see text for details]{Pillepich_2019} are shown in grey squares. The fiducial measurements of this work (i.e., within a radius of $r_{\rm e}$ of \Halpha, 0.5 ckpc pixels, unweighted) are shown as black diamonds. Light green points correspond to SFR-weighted measurements with the same binning scale as the fiducial. Orange points correspond to measurements with a pixel size of 5 ckpc (i.e., 10x worse spatial resolution than the fiducial). Purple diamonds correspond to measurements with both SFR-weighting and a pixel size of 5 ckpc. The black, grey and purple points are shown in larger sizes for emphasis (same as in Fig.~\ref{fig:kin_vs_z}). Intrinsic measurements of TNG50 galaxy kinematics can vary by up to a factor of $\lesssim$2 at fixed redshift, depending on e.g. choices of aperture, spatial binning and SFR-weighting.}
        \label{fig:TNGvandsigma_vs_z}
    \end{figure}
    
    First, for each galaxy, we calculate the reduced quadrupole moment tensor of the gas mass distribution, which describes the deviations of the shape of the galaxy from a sphere. We then compute the eigenvalues of the mass tensor, i.e., the longest, middle, and shortest axis lengths ($A,B,C$), respectively \citep{Genel_2015}. We use these axis lengths to switch to a coordinate system defined by their corresponding eigenvectors, with the galaxy 'up-vector' defined along the eigenvector corresponding to $C$. To measure the intrinsic rotation profile, we project the galaxy edge-on along the longest axis and select all star-forming gas within a virtual slit. We define the slit using a width (y-axis) of $[-r_{\rm e}/2,r_{\rm e}/2]$\footnote{Erratum: We report here an error in \citep{Pillepich_2019}, wherein they reported using a slit height of [-1/5,1/5] times the stellar half-mass radii. However, the measurements were made with a slit width of double that value.}, and a length (x-axis) and depth (line-of-sight) of $[-r_{\rm e}/2,r_{\rm e}/2]$ each in pkpc, from the centre of the {\sc SUBFIND} object (galaxy). Here, $r_{\rm e}$ is the \Sersic\ fit effective radius obtained from mock NIRCam imaging (see Sec.~\ref{sssec:sersic}). We only select gas gravitationally bound to the galaxy, and thereby ignore unbound in/out-flowing gas and nearby satellites. We then compute the unweighted mean of the line-of-sight velocities, $v_y$, in bins of 0.5 comoving kpc (ckpc) along the slit to obtain a 1D mean velocity profile as a function of distance $x$ for each galaxy. We define the absolute maximum of this profile as the galaxy's rotational velocity, $v_{\rm rot}$. We choose this definition to compare it with the velocity at $r_{\rm e}$, $v(r_{\rm e})$, inferred from observations. Although these are comparable definitions, our intrinsic measurement choice differs slightly, i.e., it is the maximum of the rotation curve within $r_{\rm e}$ rather than the measurement at $r_{\rm e}$.

    To measure the intrinsic dispersion of the galaxy, we project the galaxy face-on along the shortest axis, and select all star-forming gas within a circular aperture of $r_{\rm e}$ kpc, including also all gas cells in the line-of-sight that is bound to the galaxy. We compute the unweighted standard deviation of the line-of-sight velocities, $v_z$, in annular bins of 0.5 ckpc around the centre of the galaxy, to obtain a radial velocity dispersion profile. We define the unweighted mean of this profile as the galaxy's velocity dispersion, $\sigma$. We note that this definition excludes any ``in-plane'' dispersions and unresolved thermal motions of the gas cells \citep[see Sec.~3 of][for a detailed discussion]{Pillepich_2019}.
    
    In Fig.~\ref{fig:TNGvandsigma_vs_z}, we show the evolution of $v$, $\sigma$, and $v/\sigma$ as a function of redshift, for galaxies in the stellar mass bin of $M_*=[10^9-10^{10}]\,\Msun$. Our fiducial choices described above are shown in black. The measurements of \citet{Pillepich_2019} are shown in grey. We also show values for \TNGvre\ and \TNGsigma\ for alternate measurement choices in different colours. In particular, we vary the weighting scheme (both unweighted and SFR-weighted) and the spatial binning scale (0.5,5 ckpc) for both quantities. As reported by \citet{Pillepich_2019}, $\sigma$ increases with redshift. We find that weighting by SFR does not change the rotational velocities much, but reduces the dispersions by $\sim20\,\kms$, thereby increasing $v/\sigma$ by about one dex on average. In contrast, increasing the spatial scales for binning (i.e., poorer resolution) does not affect $\sigma$ much, but reduces the velocities by $\sim 30\,\kms$ on average and significantly impacts its scatter at fixed redshift. In this case, $v/\sigma$ also reduces by one dex on average.

    When considering measurements with both SFR-weighting and 10x larger spatial scales, the average $v$, $\sigma$, and $v/\sigma$ decrease, while the scatter extends to lower values. The higher incidence of cases with low $v$ is due to the maximum of the rotation curve not being resolved in measurements with 10 times larger spatial bins (see 16th percentiles of orange and purple markers).
    
    We find that our intrinsic measurements of $v$ and $\sigma$ differ from those of \citet{Pillepich_2019} (shown in grey) due to the following: Firstly, \citet{Pillepich_2019} measure the rotational velocity maximum within twice the stellar half-mass radius ($2\times r_*$). This aperture tends to be larger than $r_e$, especially for $M_*\leq10^{10}\,\Msun$ galaxies. This implies that the measurements of this work (shown in black) may not probe the velocity maximum in these galaxies. Secondly, \citet{Pillepich_2019} measure $\sigma$ in a slit along the galaxy's shortest axis but ignore contributions from the regions within $r_*$. We use a circular aperture for this work, as azimuthal variations due to turbulent, star-forming clumps can bias measurements of $\sigma$. Finally, \citet{Pillepich_2019} exclude the central region of the galaxy where the velocity dispersions are typically higher, causing their measurements to be lower than this work (see top right panel of Fig.~\ref{fig:TNGvandsigma_vs_z}).
    
    We emphasise that these ``intrinsic" measurements are not an absolute truth and are subject to several choices. For this work, we measure \TNGvre\ and \TNGsigma\ as described above, considering them conceptually closest to observationally derived kinematics (e.g., unweighted measurement for all star-forming gas within $r_{\rm e}$). Yet, differences across measures highlight the intrinsic uncertainties in measuring $v$ and $\sigma$, purely due to choices in aperture, weighting, and spatial binning. Observationally, these choices translate into instrumental limitations and imply that NIRSpec observations may underestimate $v$ and $\sigma$ due to spatial resolution and emission weighting, respectively.
    
    \subsection{Mock JWST observations of TNG50 galaxies}
        \label{subsec:mock}

        \begin{figure*}
            \centering
            \includegraphics[width=1\textwidth]{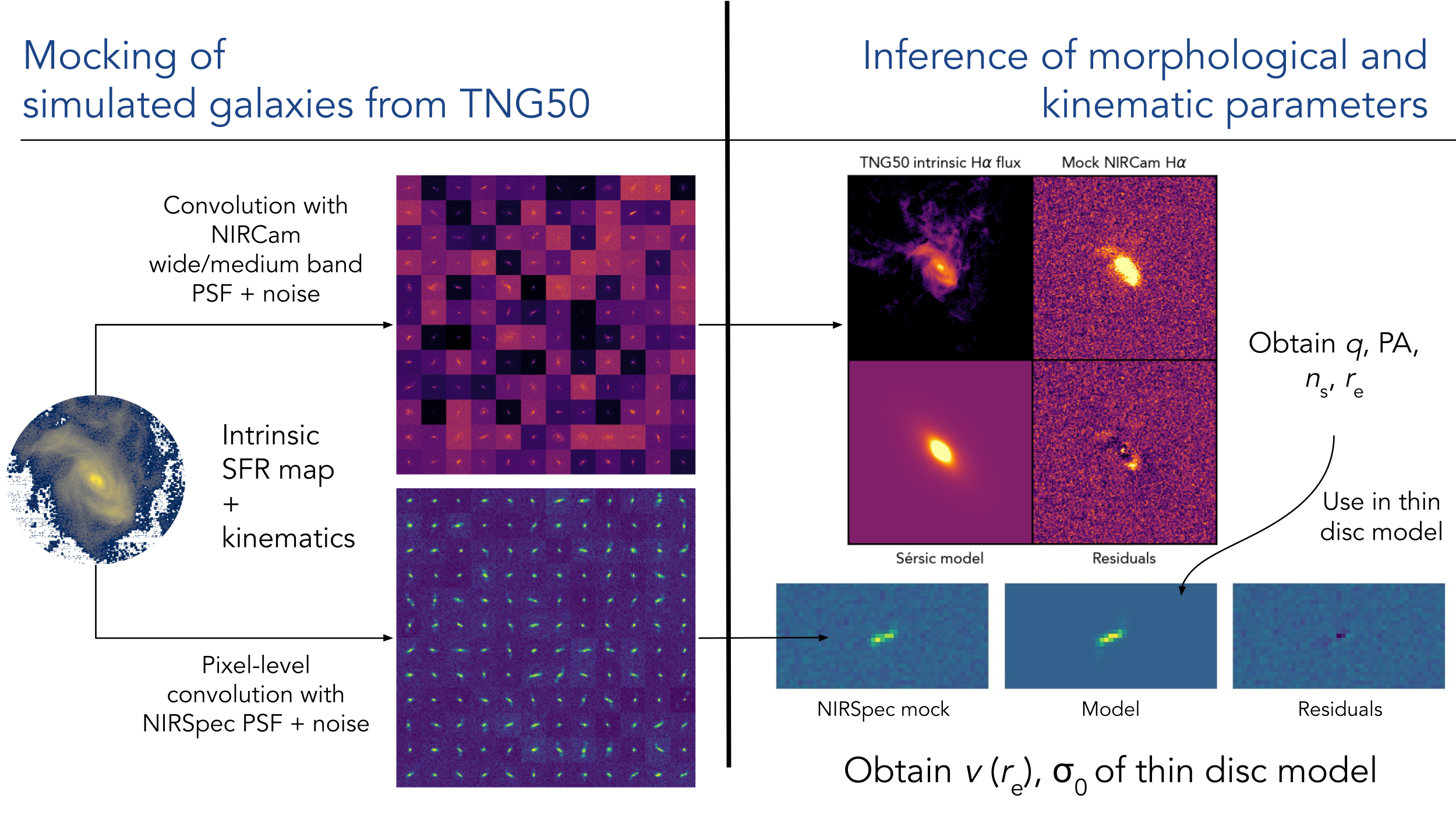}
            \caption{Schematic of the mocking procedure (Sec.~\ref{subsec:mock}), and inference of the morphological and kinematic properties (Sec.~\ref{subsec:inference}) adopted throughout this paper and shown here for a single example object. The positions, velocities, and SFRs of the star-forming gas in each simulated galaxy from TNG50 (left) is used to create intrinsic images or cubes. The images (cubes) are then convolved with a NIRCam (NIRSpec) PSF and noise is added (see 12$\times$12 collages). The mock images are then fit with \pysersic\ to obtain morphological parameters such as $q$, PA, $n_{\rm s}$ and $r_{\rm e}$. These parameters are then used in conjunction with kinematic forward models (assuming a thin, rotating disc) to infer the kinematics of the galaxies. We thus obtain the rotational velocity at $r_{\rm e}$ \msavre\ and the velocity dispersion \msasigma\ for each galaxy.}
            \label{fig:pipeline_schematic}
        \end{figure*}

        In this subsection, we describe the procedure for making mocks of NIRCam photometric images and NIRSpec/MSA spectra. Predicting \Halpha\ emission from simulations is challenging, and requires full radiative transfer post-processing of the gas (not done in TNG50 explicitly). Since our goal is to test the accuracy of kinematic inference rather than the sub-grid physics employed for TNG50, we undertake a simplified approach to obtain the \Halpha\ emission in star-forming regions. We hence use the calibration of \citet{Kennicutt_1998} to associate the instantaneous SFR of star-forming gas in TNG50 to \Halpha\ luminosities as follows:

        \begin{equation}
            \mathrm{SFR\,[\Msun\,yr^{-1}]\,=\,7.9\times10^{-42}\,L(H\alpha)\,[erg\,s^{-1}]}.
            \label{eq:Kennicutt}
        \end{equation}
        
        We assume that the relation between SFR and \Halpha\ luminosity holds at the $\sim$sub-kpc scales probed by most telescopes at cosmic noon \citep[see][for remarks on \Halpha\ as a tracer of SFR in the local and high-redshift Universe]{Peters_2017,Hu_2024,Kramarenko_2025}. Furthermore, we neglect the effects of dust attenuation on the \Halpha\ luminosity and defer the effects of dust to future work.
        
        We analyse each galaxy independently, i.e., we only consider star-forming gas associated with the galaxy as identified by {\sc SUBFIND}. We thus neglect contributions from gas in the halo or nearby companions. Below, we describe the mocking procedure for TNG50 galaxies in two steps. First, we generate mock \Halpha\ images to infer morphological quantities. Second, we generate mock \Halpha\ spectra to infer parameters that include the kinematics of the galaxy.
    
        \subsubsection{NIRCam mock images}
            \label{sssec:nircam}

        Although a dedicated image simulation tool for NIRCam does exist (\texttt{MIRAGE}; \citealt{Hilbert_2019}), and was used by \citet{Costantin_2023} to create NIRCam mock images of the stellar light of TNG50 galaxies, this software was designed to test data reduction and calibration pipelines. This tool thus outputs single exposures that include a high level of data complexity (e.g., cosmic rays, correlated noise). Consequently, constructing mock images comparable to the drizzled images typically used in observational studies would require a computationally intensive reduction process for each simulated galaxy. However, in this work, the mock images serve only to inform the kinematic inference (see Sec.~\ref{subsec:inference}) by predetermining the approximate emission line morphology, which does not require the sophisticated image simulation features of \texttt{MIRAGE}. We therefore generate mock NIRCam images using a simpler procedure, similar to that described in \citet{deGraaff_2022}, with the key difference that we consider only \Halpha\ emission from star-forming gas, and JWST rather than ground-based imaging.

        For each TNG50 galaxy in our sample, we create a random projection of the star-forming gas in 2D by simply projecting along the z axis of the simulation box. We centre the galaxy image using the position of the {\sc SUBFIND} object, defined as the location of the particle with the minimum gravitational potential energy, expressed in physical kiloparsecs (pkpc). The spatial extent is defined as either the NIRSpec slit length of 1.59$^{\prime\prime}$, or four times the stellar half-mass radius, whichever is larger (in pkpc) at the target redshift. We include all star-forming gas (belonging to the galaxy) along the line of sight. We then bin the gas cells at an oversampled angular resolution of 0.01$^{\prime\prime}$, and sum the gas SFR within each bin.
        
        Next, we compute the corresponding \Halpha\ luminosity in each bin using Eq.~\ref{eq:Kennicutt}, and convert the \Halpha\ luminosity to a flux using the luminosity distance $\mathrm{D_L}$ at the respective redshift. We assume that only \Halpha\ from the star-forming gas contributes to the flux in the NIRCam filter, neglecting for example, the stellar light, and calculate the transmission through the NIRCam filter curve using \texttt{sedpy} \citep{Johnson_2019}. We perform a point spread function (PSF) convolution of the resulting intrinsic flux image with PSF models generated using WebbPSF \citep{Perrin_2015}. We select the following NIRCam medium-band and wide-band filters corresponding to the rest-frame wavelength of \Halpha: $F200W, F277W, F335M, F460M$ for redshifts $z=2, 3, 4, 6$, respectively. We then downsample the image to an angular resolution of 0.02$^{\prime\prime}$ or 0.04$^{\prime\prime}$, for the short ($<2.25\,\mu m$) and long wavelength filters, respectively. These pixel scales are slightly better than the native sampling of NIRCam, and are commonly used in generating large image mosaics \citep[e.g.][]{Eisenstein_2023,Valentino_2023}.
        
        We adapt the procedure of \citet{deGraaff_2022} to convert these noise-free images into mock images with realistic noise properties. The fluxes are converted to counts using the typical NIRCam detector gain reported in the JWST User Documentation (JDox)\footnote{\url{https://jwst-docs.stsci.edu/jwst-near-infrared-camera/nircam-instrumentation/nircam-detector-overview/nircam-detector-performance\#gsc.tab=0}}. We consider three sources of noise: the dark current, read noise, and Poisson noise from the source itself. To estimate the former two, we use typical values from the JDox (with a user-specified exposure time and number of exposures) and compute the Poisson noise resulting from the total photoelectron counts. We consider an exposure time of 1375 seconds and a total of 26 exposures, similar to the deepest observations obtained with JADES \citep{Eisenstein_2023}. After subtracting the dark current, we convert the image back to physical units (i.e., $\mu$Jy) to obtain the final mock NIRCam image.
        
        We note that we do not consider other sources of background noise (e.g. zodiacal light), which implies that we underestimate the true noise level. Therefore, although we used the exposure setup of the JADES program in the GOODS-S field, in practice we find that the point source depth of the mock images ($\sim1-2$\,nJy) is instead comparable to the deeper imaging of the JADES Origins Field \citep{Eisenstein_2025}. However, this is balanced by the fact that we only consider \Halpha\ emission and do not include stellar continuum emission or nebular emission from other lines in the NIRCam filters, which would substantially boost the observed flux for typical galaxies. The overall noise properties of the mock images are thus still comparable to deep imaging surveys such as JADES and MEGAScience \citep{Eisenstein_2023,Suess_2024}.
        
        \subsubsection{NIRSpec mock spectra}
            \label{sssec:nirspec}

        To extract kinematic properties of TNG50 galaxies, we generate mock spectra by first creating 3-dimensional spectral flux cubes $F(x,y,\lambda)$ for each galaxy in our sample. Similar to the previous section, each galaxy is projected along one of the box axes (z-direction). This results in a random alignment of the slit with respect to the kinematic projected major axis. We fix the spatial extent to be twice the half-gas mass radius of the galaxy, measured from the centre of the {\sc SUBFIND} object. This spatial selection typically extends well beyond the range of the star-forming regions of the galaxy. We set the spatial bins in pkpc based on an angular resolution of 0.024$^{\prime\prime}$ at the specified redshift, chosen to match the JWST PSF grid and ensure it is well-sampled.
        
        We use the line of sight ($v_z$) velocities of each star-forming gas cell, subtracted from the peculiar velocity of the galaxy as defined by {\sc SUBFIND} to compute bins in wavelength space, i.e., $\mathrm{\Delta\lambda=\lambda_0\times v/c}$, where $\lambda_0$ is the observed-frame wavelength of \Halpha\ at the respective redshift. We ignore contributions from the thermal state of individual gas cells to the velocity profile of the \Halpha\ emission line. We set a velocity resolution of $10\,\kms$ to create a high-resolution flux cube. Following Eq.~\ref{eq:Kennicutt}, we compute the \Halpha\ spectral flux for each $(x,y,\lambda)$ bin using the instantaneous gas SFR and luminosity distance at the respective redshift, in units of $\mathrm{erg\,s^{-1}\,cm^{-2}}$\AA.

        We adapt \msafit\ \citep{deGraaff_2024} to propagate the \Halpha\ flux cubes through the NIRSpec instrument to produce 2-dimensional NIRSpec/MSA spectra. We refer the reader to \citet{deGraaff_2024} for details on the package, and summarise the procedure in the following. First, we choose the appropriate filter and high-resolution disperser combination for the respective redshift, i.e., $F170LP/G235H$ for $z=2,3$ and $F290LP/G395H$ for $z=4,6$. We consider the slit long axis to be along the vertical direction (see Fig.~\ref{fig:collage}). We convolve each 3D bin of the spectral cube with its unique PSF from the NIRSpec PSF libraries, determined by the Filter-Disperser combination, while also accounting for bar shadows. The convolved spectrum is then placed on a mock detector using the trace libraries, and then downsampled to the true detector pixel size and convolved with the inter-pixel capacitance kernel \citep{deGraaff_2024}.        
        
        Finally, we apply random Gaussian noise, described by a Gaussian centred at zero flux with a non-zero standard deviation, $\sigma_{\rm noise}$. This is computed as the maximum of $F_\mathrm{N}/\mathrm{(SNR}\times\sqrt{N})$ and $4\times10^{-21}\,\mathrm{erg/s/cm^2/}$\AA, thereby providing a signal-to-noise ceiling comparable to the deepest high-resolution JWST NIRSpec/MSA spectra that currently exist in the JWST archive (GO-1871; PI: Chisholm; see e.g. \citealt{Chisholm_2024}). Since we do not consider Poisson noise from the source itself (due to the fact that the noise properties of the NIRSpec instrument are highly challenging to capture in a simple model, see e.g. \citealt{Jakobsen_2022}), we apply a maximum $(S/N)$ of 50 to avoid unrealistically high $(S/N)$ for very bright sources. $F_\mathrm{N}$ is the flux in the brightest $\mathrm{N}$ pixels of the 2D spectrum, which we fix to be 12, chosen as this number represents the typical number of $(S/N)>1$ pixels in a mock spectrum. We note that the observing setup here is deliberately chosen to be close to ideal, as the key aim of this work is to test the kinematic inference methodology itself. In addition to choosing a deep exposure time (i.e. high $(S/N)$), we therefore neglect hot pixels and cosmic rays, and assume the slit is centred perfectly. We apply the mocking procedure described in Sec.~\ref{subsec:mock} to all galaxies in our sample. A schematic of the procedure for a single galaxy and a representative collage of 144 (12x12) mock NIRCam images and NIRSpec spectra are shown in Fig.~\ref{fig:pipeline_schematic}.
        
    \subsection{Inference of morphological and kinematic properties}
        \label{subsec:inference}

        As outlined in Sec.~\ref{sec:intro}, the goal of this work is to infer the kinematics of high-redshift galaxies, given a kinematic model described by a thin rotating disc. However, the imprint of different morphological and kinematic properties ($v$ and $\sigma$) on the 2D spectra can be degenerate (for e.g., due to possible misalignment of the slit with respect to the kinematic major axis). We thus attempt to make strong constraints on the morphological information first, namely, the $q$ (axis ratio), $r_{\rm e}$ (effective radius), PA (position angle), and $n_{\rm s}$ (\Sersic\ index), before fitting for the kinematics of the galaxy. Although the morphology of the stellar light is more commonly probed observationally, the \Halpha-emitting gas and stellar light have similar morphologies (albeit slightly larger \Halpha\ sizes) for star-forming galaxies at $z\sim1$ \citep{Nelson_2016}. We hence use the highly constrained priors for the \Halpha\ morphology in the kinematic fitting procedure and defer an exploration of obtaining morphologies from the stellar light to future work. We describe these steps as follows.
        
        \subsubsection{\Sersic\ fitting}
            \label{sssec:sersic}

        \begin{table}
            \centering
            \begin{tabular}{c c c}
                \hline
               Parameter  & Prior type & Prior (low,high) or \{$\mu$,$\sigma_{\rm prior}$\} \\
               \hline
               Position Angle: PA & Uniform & $[0,\pi)$\\
               \Sersic\ index: $n_{\rm s}$ & Uniform & $[0.65,8)$\\
               Ellipticity: $\epsilon$ & Uniform & $[0,0.9]$\\
               Effective radius: $r_{\rm e}$ & Gaussian & \{0.2$^{\prime\prime}$,0.2$^{\prime\prime}$\}\\
               Flux & Gaussian & \{Image flux, Image flux/2\}\\
               Offsets from center: $rx\,\&\,ry$ & Gaussian & \{Image center, 3 pixels\}\\
               \hline
            \end{tabular}
            \caption{Priors for \pysersic\ fitting.}
            \label{tab:sersic_priors}
        \end{table}        

        For each NIRCam mock as derived using the steps in Sec.~\ref{sssec:nircam}, we compute the ``observed'' $(S/N)$ in a circular aperture around the image centre, defined by a radius of 0.12$^{\prime\prime}$. We select mock sources with $(S/N)>5$ and fit each selected image as a single source with the Python package \pysersic\footnote{\href{https://pysersic.readthedocs.io/en/latest/}{https://pysersic.readthedocs.io/en/latest/}}\ \citep{Pasha_2023}.

        \begin{figure}
            \centering
            \includegraphics[width=0.5\textwidth]{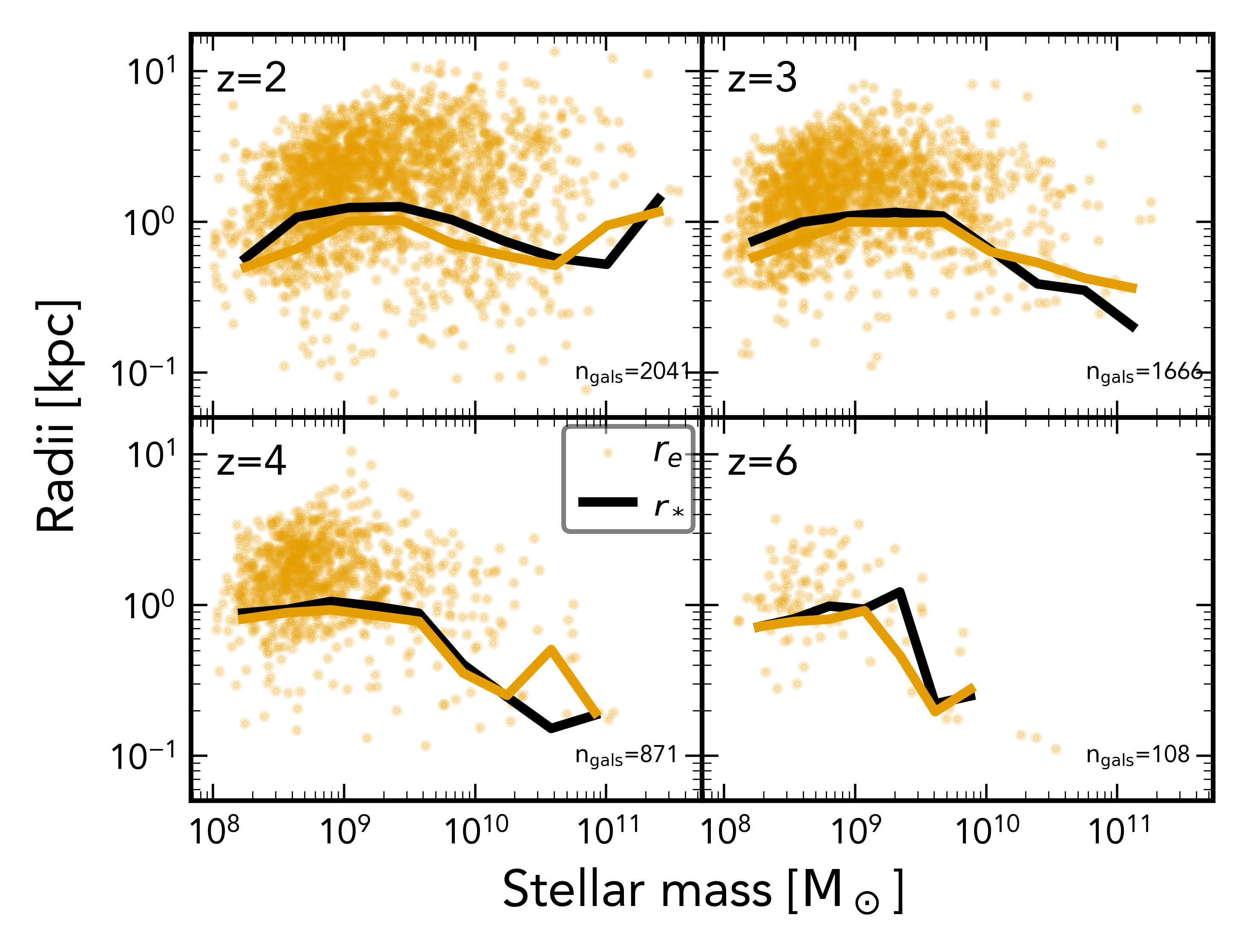}
            \caption{Sample distribution of mock \Halpha\ image sizes ($r_{\rm e}$) versus galaxy stellar mass ($\Mstar$) for TNG50 galaxies with NIRCam $(S/N)>10$, shown as orange points. The running averages for \Halpha\ effective radii and 3D stellar half-mass radii are indicated in orange and black, respectively. The number of galaxies and the redshift of the snapshot are annotated in each panel. The 3D stellar half-mass radii and 2D \Halpha\ effective radii agree well with each other on average.}
            \label{fig:size_mass}
        \end{figure}

        Following the setup of \citet{Miller_2025} to estimate rest-frame optical sizes, we set uniform priors for the PA, $n_{\rm s}$, and ellipticity ($\epsilon=1-q$) of $[0,\pi)$ radians, $[0.65,8)$, and $[0,0.9]$, respectively. We set Gaussian priors for the $r_{\rm e}$, flux, and $rx\,\&\,ry$ (position offsets of the \Sersic\ profile centre from the image centre). We choose initial guesses ($\mu$, i.e., mean of the Gaussian) of 0.2$^{\prime\prime}$, the total image flux, and the image centre for them, respectively. The widths of the Gaussians ($\sigma_{\rm prior}$) are 0.2$^{\prime\prime}$, half the image flux, and 3 pixels, respectively, with the $r_{\rm e}$ and flux forced to be non-negative. The chosen priors are summarized in Table~\ref{tab:sersic_priors}.

        We fit a 2D \Sersic\ model for each mock image using \pysersic\ with the set priors. \pysersic\ uses the NUTS Markov Chain Monte Carlo (MCMC) algorithm \citep{NUTS} to infer the posterior distribution of the morphological parameters. An example fit is shown in Fig.~\ref{fig:pipeline_schematic} (top right). Galaxies with $(S/N)<$10 often have poorly constrained posteriors, and we thus discard any such mock galaxies from further analysis.
        
\begin{figure}
            \centering
            \includegraphics[width=\columnwidth]{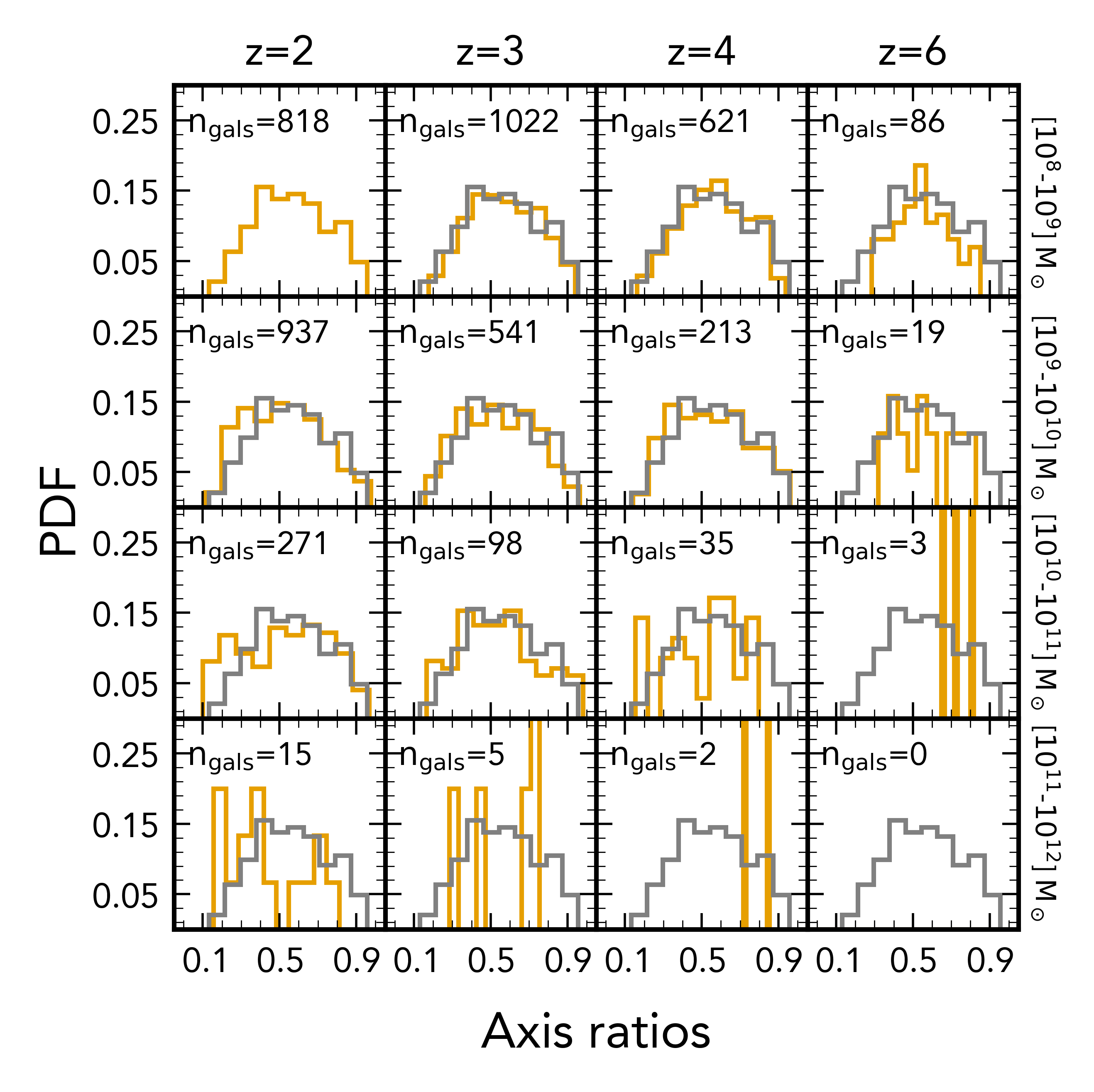}
            \caption{Probability distribution of axis ratios ($q$) inferred from the mock NIRCam imaging of TNG50 galaxies with stellar masses of $10^8-10^{11.5}\,\Msun$ and NIRCam $(S/N)>10$. The columns describe the redshift of the galaxy while the rows show the various bins in stellar mass (see annotations on the right). The gray line indicates the axis ratio distribution of the top left panel ($z=2$, $M_*=[10^8-10^9)\,\Msun$) for comparison. More massive galaxies at lower redshifts exhibit the smallest axis ratios, down to values of 0.1.}
            \label{fig:axis_ratios_pdf}
        \end{figure}

        Fig.~\ref{fig:size_mass} shows the distribution of galaxy effective radii ($r_{\rm e}$), galaxy 3D stellar half-mass radii (averaged), and stellar masses for galaxies with $(S/N)>10$. The number of galaxies selected decreases with stellar mass and redshift, as expected. Further, the effective radii $r_{\rm e}$ of central galaxies do not change much with galaxy stellar mass at $z=2$, while they begin to decrease for $z\geq3$. However, we note that for the most massive galaxies, the lack of dust post-processing in our mocking procedure makes dense starbursts to have extremely compact $r_{\rm e}$. In contrast, simulations with appropriate prescriptions for dust (either on-the-fly or via post-processing) find rest-frame optical sizes to increase with stellar mass \citep{Ma_2018,Roper_2022}, owing to centrally concentrated dust distributions \citep{Popping_2022}. Yet, JWST has revealed discrepancies in the steepness of the size-mass relation for $M_*>10^{10}\,\Msun$ galaxies \citep{Allen_2025} \citep[but see also][]{McClymont_2025}. Fig.~\ref{fig:axis_ratios_pdf} shows the probability distribution of axis ratios binned in stellar mass for each snapshot. The axis ratios are uniformly spread, with a slight overall increase with increasing redshift and decreasing stellar mass.

        We use the \pysersic\ posterior distribution in the following section to add strong constraints to the morphology while fitting the kinematics of the galaxy.
            
        \subsubsection{Kinematic modelling}
            \label{sssec:kin_model}

        For each galaxy with high $(S/N)$ ($>10$), we use \msafit\ to fit the 2D mock spectrum (see Sec.~\ref{sssec:nirspec}) using a rotating thin disc model with an arctangent velocity profile and constant velocity dispersion. The procedure is presented in \citet{deGraaff_2024} and we provide key aspects and modifications below.

        The thin disc model is defined by 11 parameters --- ($\lambda_0$, $F$, $rx$, $ry$, $r_{\rm e}$, $r_{\rm t}$, $n_{\rm s}$, $q$, PA, $v_a$, $\sigma_0$), whose descriptions are in Table~\ref{tab:kin_model_params}. We model the emission surface brightness distribution of the galaxy as a \Sersic\ profile \citep{Sersic_1968} and the velocity profile as $v(r)=\frac{2}{\pi}v_a\,{\rm arctan}(r/r_{\rm t})$ \citep{Courteau_1997} with a constant velocity dispersion of $\sigma_0$. Here, $v_a$ is the asymptotic rotation curve velocity. We deliberately choose a simple geometric model in order to test the efficacy of such models in describing high-redshift galaxies. Moreover, constraining more parameters with a single spectrum of $O(10)$ pixels is challenging, motivating us to reduce complexity.

        \begin{table*}
            \centering
            \begin{tabular}{c c c c}
                \hline
               Parameter  & Description & Units & Vary?\\
               \hline
               $\lambda_0$ & \Halpha\ line centroid wavelength & \AA & \checkmark \\
               $F$ & Integrated flux & $\mathrm{erg\,s^{-1}\,cm^{-2}}$ & \checkmark \\
               $rx$ & Position offset in the dispersion direction & arcseconds & $\times$ \\
               $ry$ & Position offset in the cross-dispersion direction & arcseconds & $\times$ \\
               $r_{\rm e}$ & Effective radius & arcseconds & \checkmark\\
               $r_{\rm t}/r_{\rm e}$ & Ratio of turnover radius, i.e., radius at which $v(r)=\frac{v_a}{2}$, to $r_{\rm e}$ & - & $\times$\\
               $n_{\rm s}$ & \Sersic\ index & - & $\times$\\
               $q$ & Axis ratio & - & \checkmark\\
               PA & Position angle & - & \checkmark\\
               $v_a$ & Asymptotic velocity of the rotation curve & $\kms$ & \checkmark\\
               $\sigma_0$ & Constant velocity dispersion & $\kms$ & \checkmark\\
               \hline
            \end{tabular}
            \caption{Thin disc model parameters in \msafit\ considered in this work. Of these parameters, we mostly focus on the velocity dispersion \msasigma, and on the rotational velocity \msavre\ rather than $v_a$, with the former being the velocity of the rotation curve at $r_{\rm e}$.}
            \label{tab:kin_model_params}
        \end{table*}

    \begin{figure*}
        \centering
        \includegraphics[width=0.75\textwidth]{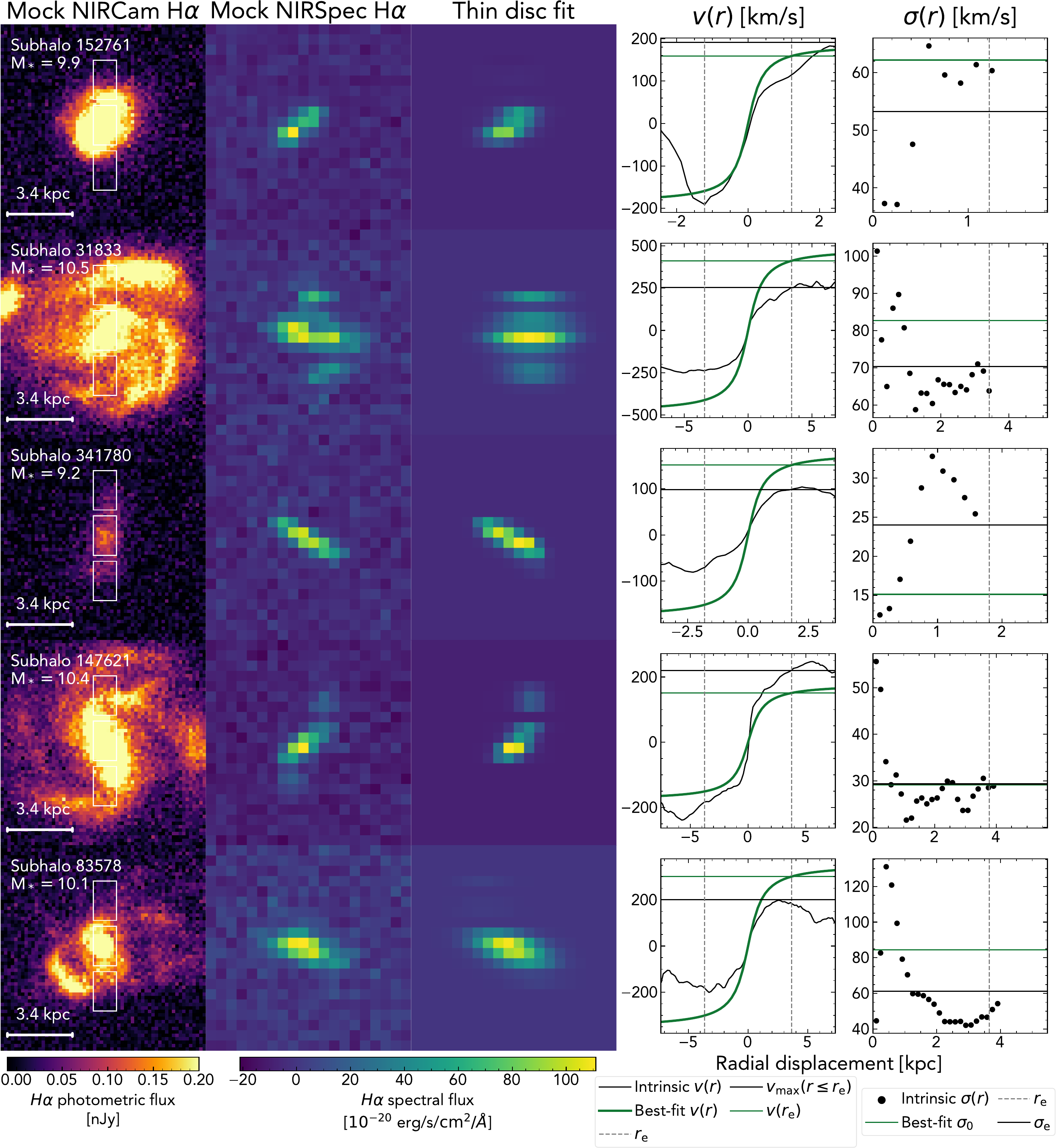}
        \caption{Collage of fit \Halpha\ photometry and kinematics for aligned and resolved $z=2$ TNG50 galaxies with good kinematic fits (i.e., low $\chi^2$ - see Sec.~\ref{sssec:kin_model}). The five columns correspond to NIRCam mock \Halpha\ images, NIRSpec mock \Halpha\ spectra, thin disc kinematic model fit to the spectra, intrinsic and best-fit rotational velocity profiles, and intrinsic and best-fit velocity dispersion profiles, respectively. The stellar masses and scales are indicated in the first column of each TNG50 galaxy. The NIRSpec/MSA mock slit is placed vertically in each case as indicated by the white rectangles. The thin disc model successfully reproduces intrinsic quantities in a variety of systems.}
        \label{fig:collage}
    \end{figure*}

        We create 2D forward models (similar to the mocks in Sec.~\ref{sssec:nirspec}) of such thin discs, and use a reduced $\chi^2$ minimization to fit for the kinematics of the galaxy. For the fit, we use the {\tt min\_optimizer} within \msafit, a wrapper for the {\tt scipy.optimize.minimize} module \citep{SciPy}. This method does not give a full posterior distribution, but only one best-fit value for the parameters listed in Table~\ref{tab:kin_model_params}. We add strong constraints to the morphology based on the \Sersic\ fits from Sec.~\ref{sssec:sersic}, allowing minimal room for change. This allows us to explore a broad parameter space for $v_a$ and $\sigma_0$, which describe the kinematics of the fit model.
        
        We fix the inclination of the disc using the observed axis ratio $q$ as $i=\mathrm{arccos}(q)\times\mathrm{\frac{180}{\pi}}$, and assume $r_{\rm t}=r_{\rm e}/4$, i.e., the value expected for a thin exponential disc \citep{Mo_1998}. We fix the position offsets ($rx$,$ry$) and $n_{\rm s}$ using the median values of the \Sersic\ fit posterior. For $r_{\rm e}$, $q$, and PA, we use half the span between the 1st and 99th percentiles of the posterior distribution from \pysersic\ as the allowed range for the fit defined around the median. We thus allow for minor variations in the morphological parameters.
        
        We set the initial guess for the flux to be the sum of the 3D mock cube with an allowed range between zero and thrice the flux of the cube. The initial guess for $\lambda_0$ is set to the rest-frame \Halpha\ wavelength and can vary by 5 \AA. We also allow the 2D model to have a velocity range beyond the $\lambda$ edges of the mock spectrum. We set initial guesses for $v_a$ and $\sigma_0$ to be $0-50\,\kms$, with allowed ranges of $[-500,500]\,\kms$ and $[10,200]\,\kms$, respectively. For systems whose fit $v_a$ or \msasigma\ differ by less than $2\,\kms$ from the bounds, we re-fit the spectra with a broader range of $[-1000,1000]\,\kms$ and $[10,500]\,\kms$ for $v_a$ and \msasigma, respectively.

        We are interested in modelling the rotation profile of the mock observations (reflecting the effects of the full mass distribution of the simulated galaxies) as smooth, thin rotating discs. However, the mock observations use \Halpha\ as the kinematic tracer, which has a non-smooth spatial distribution with clumps and voids. To mitigate the effect of these spatial light variations on the recovery of kinematic information, we use a non-parametric, row-wise flux scaling $S(y)$ in the kinematic modelling, following \citet{Price_2020}. This factor enables the model intensity profile to match better the mock spectrum and its non-smooth, non-\Sersic\ intensity profile. We define $S(y)$ as 

        \begin{equation}
            S(y) = \frac{\sum_xm(y)[f(x,y)f_{\rm model}(x,y)/\sigma_f(x,y)^2]}{\sum_xm(y)[f_{\rm model}(x,y)^2/\sigma_f(x,y)^2]},
        \end{equation}

         where $x,y$ are the spectral and spatial axes, $f(x,y)$ is the 2D mock spectrum, $f_{\rm model}$ is the thin disc model spectrum, and $\sigma_f(x,y)$ is the mock spectrum uncertainty on each of the flux values $f(x,y)$. Owing to the NIRSpec/MSA being highly undersampled, we discard very noisy rows to extract information only from reliable parts of the spectrum. To this end, we define row-wise masks as $m(y)\equiv(S/N)(y)\leq1$, where $(S/N)(y)$ is the total flux in a row divided by the total error in a row added in quadrature. We then only fit contiguous, unmasked rows that are not affected by the NIRSpec bar shadow. We investigated differences in fit residuals for higher or lower thresholds for the $(S/N)(y)$ masking, while being agnostic to how well the kinematics are recovered. We found that our fiducial choice performs best, as it retains most information in the spectrum for a variety of sources while avoiding over- or under-fitting.

         To offset the amplification of bright rows in the thin disc model due to fainter parts of the intensity profile or the flux scaling $S(y)$, we up(down)-weight the low (high) $(S/N)$ unmasked rows by defining weights as $w_y=[(S/N)(y)]^{-1}$. Dimmer yet reliable parts of the spectrum are thus more constraining on the fit quality, allowing us to probe $v$ and $\sigma$ to larger radii, particularly for low stellar masses. We explored other choices for the weighting such as $w_y=\sqrt{[(S/N)(y)]}$, and found minimal differences between the schemes (see Fig.~\ref{fig:SNRwt_comparison}). Finally, we use the fixed and free parameters to fit the 2D mock spectrum following a weighted $\chi^2$ method \citep[see section A.2 of][]{Price_2016}. The goodness-of-fit criterion is defined as

        \begin{equation}
            \chi^2 = \sum_{x,y}w_y\times\left[m(y)\frac{f(x,y)-S(y)f_{\rm model}(x,y)}{\sigma_f(x,y)}\right]^2.
        \end{equation}

        where $m(y)$ is to mask rows in spatial coordinates, $w_y$ are row weights, and $S(y)$ is the row-wise flux scaling between the model and mock spectrum.

        Our model successfully fits mock spectra for hundreds of galaxies. An outline of the entire mock and inference pipeline is shown in Fig.~\ref{fig:pipeline_schematic}. We successfully obtain rotational velocities and velocity dispersions for intrinsically smooth discs, and even for clumpy systems (see Fig.~\ref{fig:collage}). However, despite the flexibility in the morphology, deviations in the kinematic profiles due to inflows and outflows are not captured by the modelling, and may strongly bias the velocity profiles. Moreover, the model may not describe well the kinematics of nearly unresolved sources. We discuss the efficacy of our methodology and possible biases in Sec.~\ref{sec:inference_results} and~\ref{sec:discussion}.

\section{Results on best-fit galaxy kinematics}
    \label{sec:inference_results}

    \subsection{Overview of the kinematic fits}
        \label{subsec:inferred_sample}

    Leveraging the statistics of TNG50 and the end-to-end pipeline described in the previous sections, we measure the kinematic properties of several thousand galaxies in the redshift range of $z=2-6$ with stellar masses $M_*>10^8\,\Msun$. We first create mock NIRCam images for TNG50 galaxies with $M_*>10^8\,\Msun$ and SFR$>0$ (5575,3973,2192,455 at $z=2,3,4,6$). For galaxies with a mock NIRCam \Halpha\ $(S/N)>10$ (2631,2130,1084,124 at $z=2,3,4,6$), we create mock spectra and fit thin disc models to them to obtain $v$ and $\sigma$. We showcase the results for a selection of five $z=2$ galaxies in Fig.~\ref{fig:collage}. The columns correspond to mock NIRCam images, mock NIRSpec spectra, best-fit thin disc models, and the best-fit versus intrinsic rotational velocity and velocity dispersion profiles. We successfully recover $v$ and $\sigma$ for disc-like systems where the intrinsic velocity profiles resemble closely the best-fit \msavre\ (see top row). Furthermore, our kinematic modelling performs reasonably well for faint sources (third row) and complex systems with clumpy light distributions (second and fourth rows). We emphasise that our strong morphological constraints, combined with the weighted $\chi^2$ fitting (see Sec.~\ref{sssec:kin_model}), allow us to reproduce the morpho-kinematic complexity of a large variety of systems.

    \begin{figure}
        \centering
        \includegraphics[width=\columnwidth]{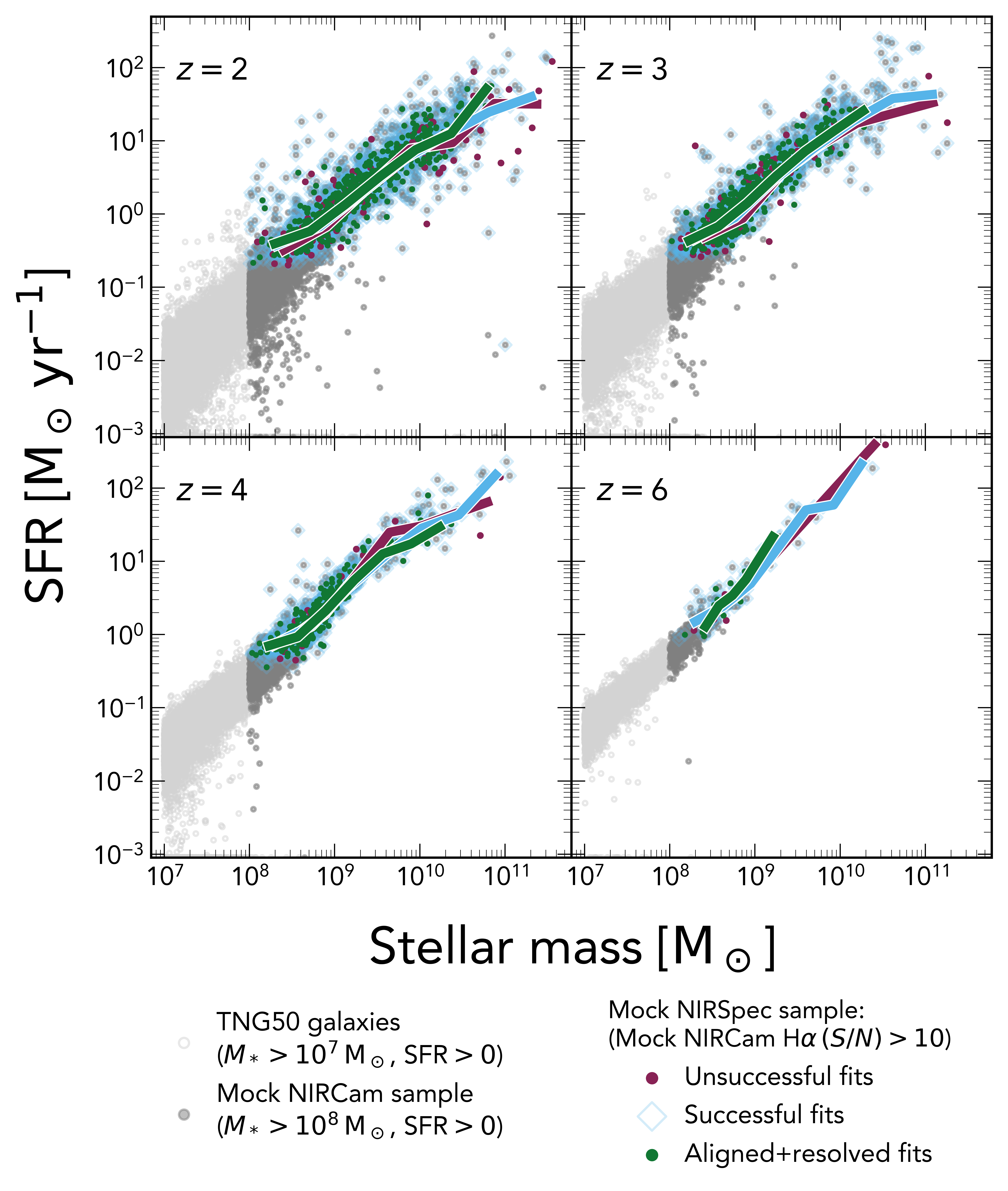}
        \caption{Star formation rate versus stellar mass of TNG50 galaxies with stellar masses of $10^8-10^{11.5}\,\Msun$ at $z=2,3,4,6$ for various subsets considered in this work (see legend). At all redshifts, galaxies with unsuccessful (magenta), successful (blue), and aligned$+$resolved (green) kinematic fits to their mock NIRSpec \Halpha\ spectra span the entire range of the star-forming main sequence beyond $M_*=10^{8.5}\,\Msun$. The most massive galaxies often do not have low $\chi^2$ for the kinematic fits due to bright clumps, thereby excluding them from the green subset. \label{fig:SFRMS}}
    \end{figure}

    \begin{figure}
        \centering
        \includegraphics[width=\columnwidth]{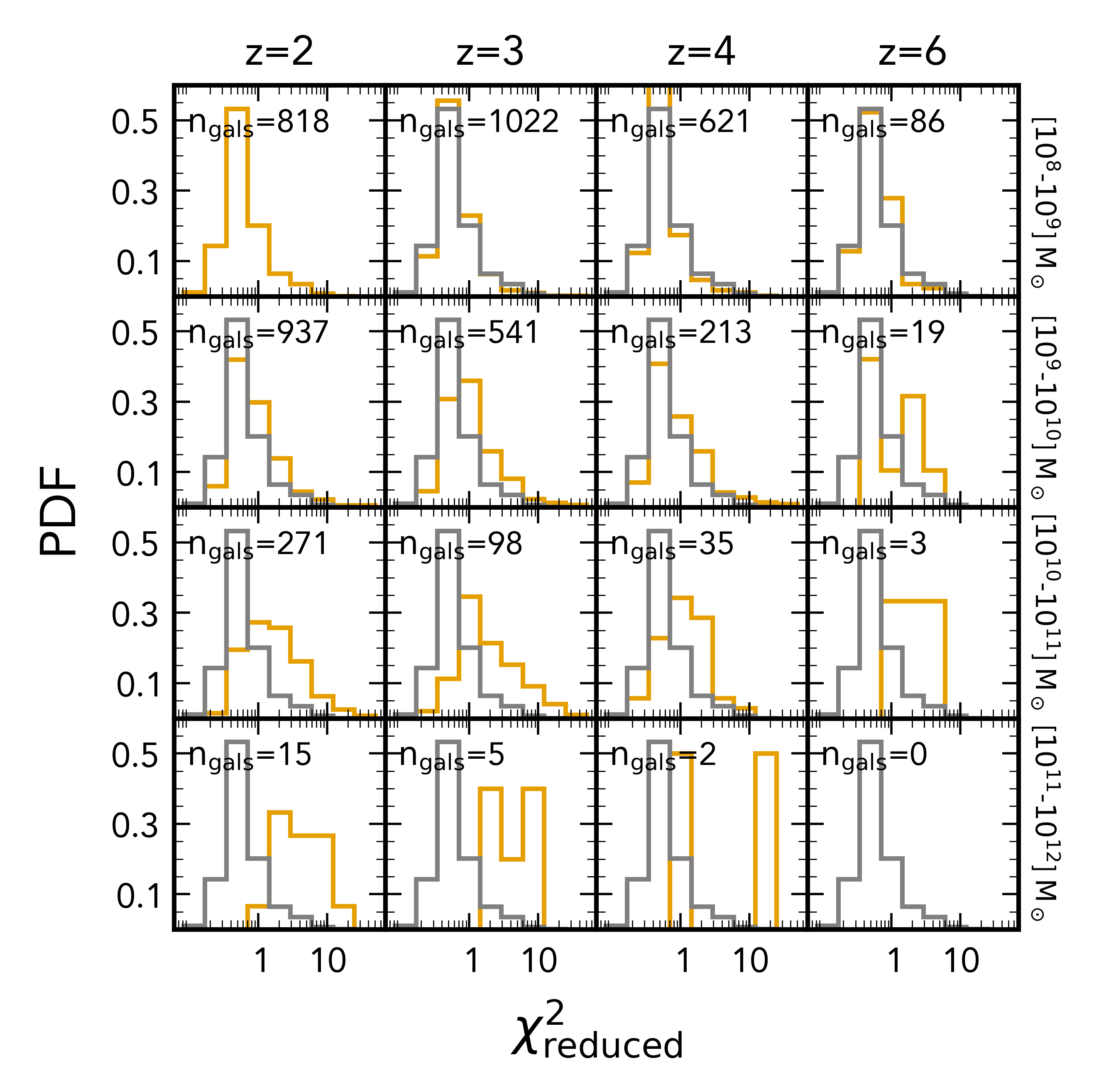}
        \caption{Quality of the kinematic fits to mock NIRSpec/MSA spectra of TNG50 galaxies with stellar masses of $10^8-10^{11.5}\,\Msun$ and NIRCam $(S/N)>10$ at $z=2,3,4,6$. We show the probability distributions of reduced $\chi^2$ for the kinematic model fits to mock spectra, in bins of redshift and galaxy stellar mass (columns and rows as in Fig.~\ref{fig:axis_ratios_pdf}). Massive galaxies at higher redshifts are fit less accurately, as they often deviate from the assumption of thin, rotating discs with \Sersic\ emission profiles.}
        \label{fig:chi2_pdf}
    \end{figure}
     
    The galaxies with NIRCam \Halpha\ $(S/N)>10$ span several orders of magnitude in star formation rates (SFR$\sim$1-100 $\Msun\,\mathrm{yr^{-1}}$) and stellar masses ($M_*=10^8-10^{11.5}$). Fig.~\ref{fig:SFRMS} shows the SFR-$M_*$ plane of TNG50 galaxies used in this work at $z=3$ (we present detailed figures for $z=2,4,6$ in Appendix.~\ref{sec:appendix_altz}). We separate the galaxies into different subsets, as labelled in the legend of the figure. These are: (i) TNG50 galaxies with a stellar mass above $10^7\,\Msun$ and SFR$>0$ (light grey), (ii) mock NIRCam sample, i.e., galaxies with a stellar mass above $10^8\,\Msun$ and SFR$>0$ (dark grey). As outlined in the Methods section, we focus on galaxies with \Halpha\ $(S/N)>10$ in the mock NIRCam imaging, hereby referred to as the mock NIRSpec sample. We further divide the galaxies with NIRSpec mocks into three subsets indicated in magenta, blue, and green for detailed analysis in the following sections: (iii) galaxies whose fit to the NIRSpec spectrum failed (151,122,71,10 at $z=2,3,4,6$), i.e., the minimization solution (via \texttt{scipy.optimize.minimize}) resulted in $v$ or $\sigma$ being too close to the bounds (magenta), (iv) galaxies that were fit successfully (1890,1544,800,98 at $z=2,3,4,6$), i.e., did not run into the bounds (blue diamonds), (v) successfully fit galaxies that have their morphological major axis aligned with the slit (PA$=\,[-45^{\circ},45^{\circ}]$, as obtained from the \Sersic\ profile fitting), are well resolved ($r_{\rm e}\geq0.1^{\prime\prime}$) and have $\chi^2_{\rm reduced}\leq2$ (590,464,213,16 at $z=2,3,4,6$) (green). We define subset (v) akin to observational slit-based studies \citep{Wisnioski_2019,Price_2020} that select on PA and size. Running medians for subsets (iii)-(v) are shown in their respective colours.

    Approximately $90-95$\% (depending on the snapshot) of our NIRSpec mocks are fit successfully. The alignment of the slit with the galaxy morphological axis is randomised as we always place the slit vertically (i.e., North-South; see Fig.~\ref{fig:collage}). Our alignment angle definition of PA$=\,[-45^{\circ},45^{\circ}]$ therefore results in only about half of the successfully fit galaxies aligning with the slit. Of these well-aligned galaxies, $\sim$80\% are also well-resolved by the JWST PSF. Further selection on the $\chi^2$ value produces our final sample of galaxies that are well described by the thin, rotating disc model (shown in green), corresponding to approximately 65\% of the aligned, successful sample (or $\sim30\%$ of all NIRSpec mocks constructed). These sources are distributed across the entire SFR-$M_*$ range along the main sequence (except for the most massive systems). Throughout, we present results for both the successful (blue) and aligned$+$resolved (green) subsets. However, we restrict our conclusions mostly to the aligned$+$resolved subset owing to its robustness.
    

    Around $\sim5-10$\% of our sample could not be fit successfully, i.e., their kinematics could not be described by a rotating thin disc model (shown in magenta). From visual inspection, we found that these galaxies can be broadly classified into three types. First, the vast majority of unsuccessful fits correspond to systems that have irregular or poorly constrained morphologies. For these sources, the \Sersic\ model fit did not provide strong constraints on PA, $r_{\rm e}$ or $q$, leading to the kinematic fit optimization incorrectly assigning dispersion features (such as broadening at a fixed row of the mock spectrum) to changes in morphological parameters. Second, galaxies with very high stellar masses were more likely to have the best-fit $v$ or $\sigma$ hit the upper bounds. Typically, these sources have steeply rising velocity profiles at the centre, (i.e. the region probed by the NIRSpec slit) which is not well captured by the arctangent model. Third, in our fitting we used a lower bound on the velocity dispersion of \msasigma$=10\,\kms$ (in order to avoid spuriously low dispersions), which for few cases of disc-like systems with intrinsically very low \TNGsigma\ implies that the fit hits the imposed lower bound and is therefore flagged as unsuccessful.

    We first assess the quality of the kinematic fits using the distribution of the weighted $\chi^2_{\rm reduced}$ in Fig.~\ref{fig:chi2_pdf}, in bins of stellar mass (rows; 1 dex per bin) and redshift (columns; $z=2,3,4,6$). The probability distribution function (PDF) of the top left panel (lowest stellar mass bin at $z=2$) is indicated in grey in the other panels for reference. The fit quality declines with redshift on average, especially for the most massive galaxies. The $\chi^2$ values of the fits are typically lower than 2 for $\Mstar<10^{10}\,\Msun$ and $z\leq4$ (note, however, that our SNR-weighted goodness-of-fit criteria in Sec.~\ref{sssec:kin_model} diminishes contributions from high SNR rows). At higher $z$, galaxies are highly clumpy and more compact, and are therefore harder to fit or are unresolved in the mock spectrum. Moreover, at higher redshifts, the intrinsic kinematic structures are more complex, and the thin disc model is therefore more likely to fail to capture this complexity. We discuss the quality of the kinematic fits and failure cases in further detail in the following subsections as well as in Sec.~\ref{sec:discussion}.

    \subsection{Fidelity of inference}
        \label{subsec:inference_fidelity}

    \begin{figure}
        \centering
        \includegraphics[width=0.45\textwidth]{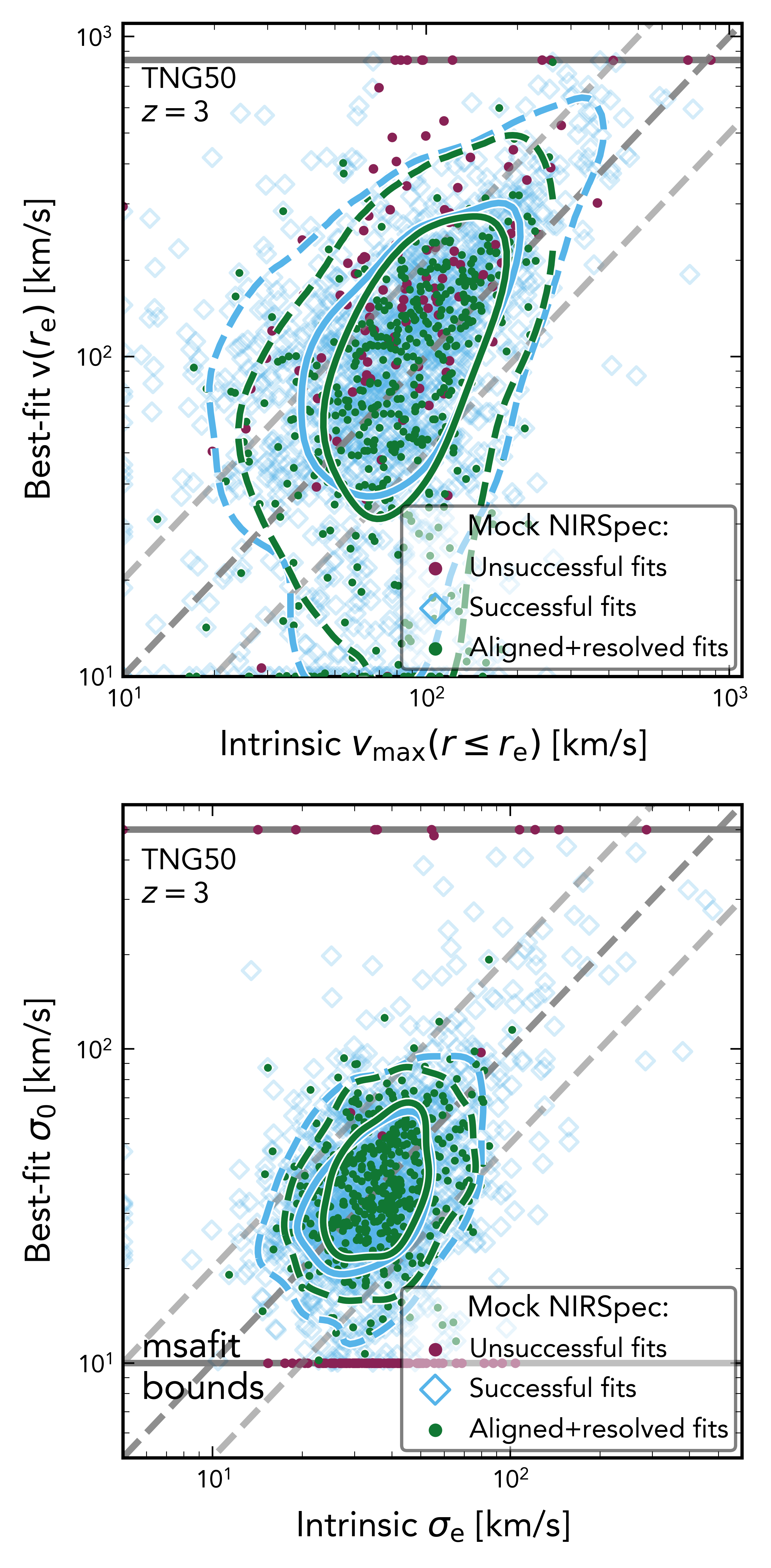}
        \caption{Best-fit \msavre\ vs intrinsic \TNGvre\ and best-fit \msasigma\ vs intrinsic \TNGsigma\ for $z=3$ TNG50 galaxies with stellar masses of $10^8-10^{11.5}\,\Msun$. The lines 1:2, 1:1, 1:1/2 are shown in dashed grey. Solid grey lines correspond to the bounds of the fitting procedure, i.e., $[-1000,1000]$ and $[10,500]\,\kms$ for $v_a$ and \msasigma, respectively. The different subsets are as in Fig.~\ref{fig:SFRMS}, here showing only unsuccessful, successful, and aligned$+$resolved fits (see legends). The solid and dashed contours of the respective colours correspond to the area within which 50\% and 84\% of the data lie. Additional redshifts are shown in Fig.~\ref{fig:bestfitvsintrinsic_altz}. Best-fit and intrinsic kinematics agree well with each other, with a wider scatter about the 1:1 line for $v$ compared to $\sigma$.} \label{fig:bestfitvsintrinsic}
    \end{figure}
    
    We compare the best-fit $v(r_{\rm e})$ and $\sigma_0$ to the respective intrinsic measurements from TNG50 (i.e., $v_{\rm max}(r\leq r_{\rm e})$ and $\sigma_{\rm e}$; see Sec.~\ref{subsec:TNG50_intrinsic}) in Fig.~\ref{fig:bestfitvsintrinsic} for the $z=3$ snapshot. The rotation velocities follow the 1:1 line, although with a large scatter above and below. Most galaxies have intrinsic $v_{\rm max}(r\leq r_{\rm e})$ values $\sim100\,\kms$, but the contours indicate cases of several under-predicted rotational velocities ($\lesssim30\,\kms$), despite having aligned and resolved kinematic fits (see green points). In contrast, both intrinsic and best-fit velocity dispersions are typically below $100\,\kms$ with minimal scatter about the line of equality, as most best-fit $\sigma_0$ values lie well within the 1:2 and 1:1/2 lines. We highlight that this scatter is therefore similarly large as the scatter in the intrinsic values themselves, which arise from differing measurement choices (see Sec.~\ref{subsec:TNG50_intrinsic}). These results hold true for galaxies at $z=2,4,6$ as well (see Fig.~\ref{fig:bestfitvsintrinsic_altz}).

    \begin{figure*}
        \includegraphics[width=0.45\textwidth]{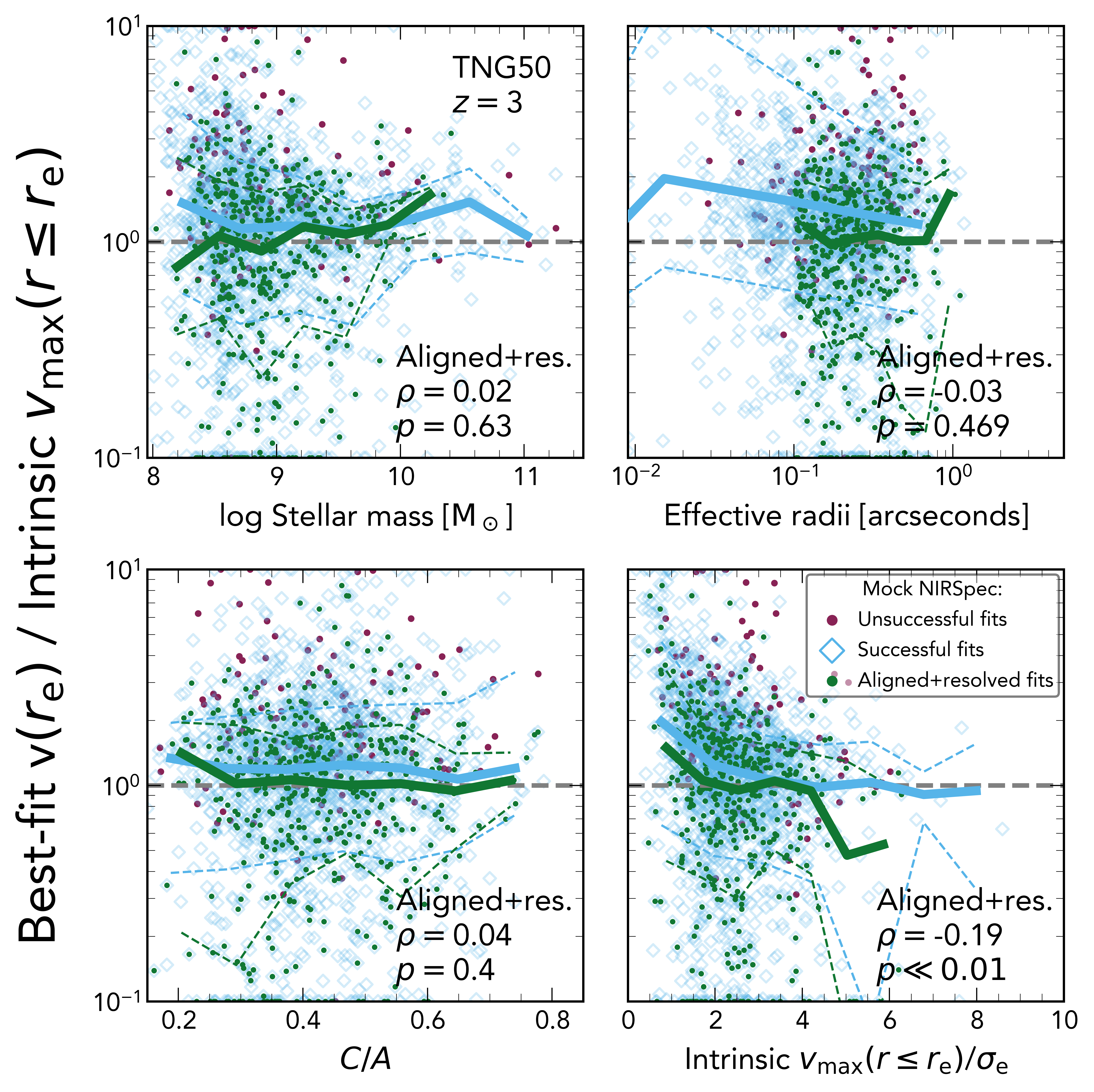}
        \includegraphics[width=0.45\textwidth]{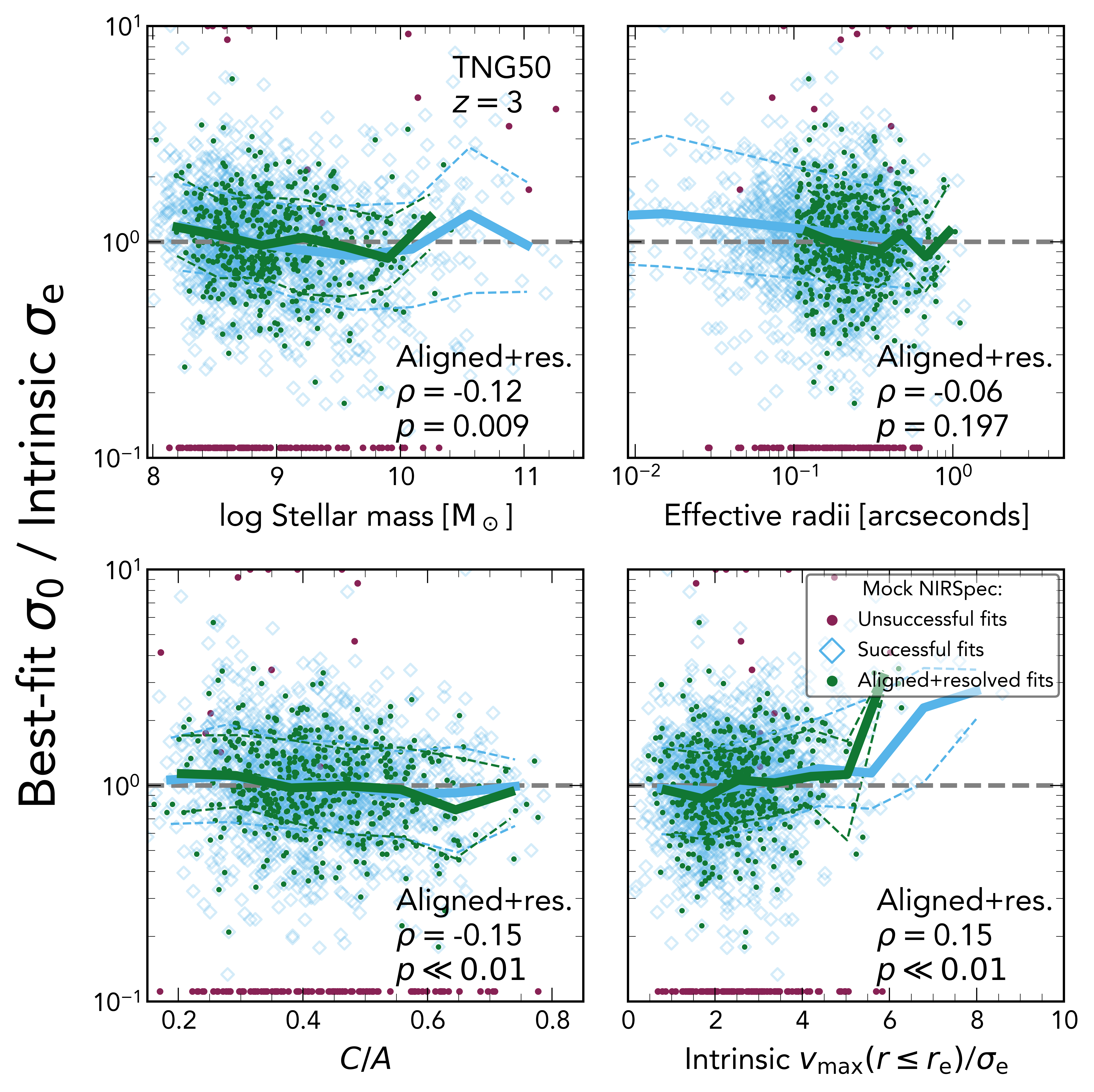}
        \caption{Ratios of best-fit to intrinsic kinematics as a function of global galaxy properties for $z=3$ TNG50 galaxies with stellar masses of $10^8-10^{11.5}\,\Msun$ (additional redshifts in Fig.~\ref{fig:bestfitbyintrinsic_vs_galprop_altz}). On the left hand side, we focus on rotational velocities, i.e. the ratio of best-fit $v(r_{\rm e})$ to intrinsic $v_{\rm max}(r\leq r_{\rm e})$, whereas on the right hand side we focus on velocity dispersions, i.e. the ratio of best-fit $\sigma_0$ to intrinsic $\sigma_e$. The four sub-panels are (i) galaxy stellar mass, (ii) \Halpha\ effective radii, (iii) Intrinsic axis ratio $C/A$, \& (iv) Intrinsic \TNGvre/\TNGsigma. The different colours correspond to subsets as in Fig.~\ref{fig:bestfitvsintrinsic} (see legend). The solid lines correspond to running medians for bins in log space. The dashed lines are 16th and 84th percentiles, respectively. Spearman rank coefficients and $p$-values for the aligned$+$resolved subset are indicated in the bottom right of each sub-panel. The dashed grey line at y-axis$=1$ is when the best-fit kinematic properties perfectly match the intrinsic measurements. The ratios of best-fit to intrinsic $v$ and $\sigma$ show nearly no dependence on stellar mass and effective radii, but are strongly (anti-)correlated with the intrinsic rotational support \TNGvre/\TNGsigma.}
        \label{fig:bestfitbyintrinsic_vs_galprop}
    \end{figure*}

    To explore possible physical causes driving the scatter in the mock-observed kinematic properties, we compare the ratio of the best-fit and intrinsic kinematics to four galaxy properties in Fig.~\ref{fig:bestfitbyintrinsic_vs_galprop}: the galaxy stellar mass, $r_{\rm e}$, the intrinsic shortest-to-longest axis ratio $C/A$, and intrinsic \TNGvre/\TNGsigma. The left set of panels shows that the best-fit \msavre\ are on average equal to the intrinsic \TNGvre, with 16th and 84th percentiles of the ratios \msavre/\TNGvre\ being 0.34 and 1.81, respectively. These statistics correspond to a scatter of a factor of $\gtrsim2$. The flat relationship with $\Mstar$ indicates that the recovery success of $v$ is independent of stellar mass. However, there is large scatter in the ratios, particularly at stellar masses $<10^9\,\Msun$, with several galaxies having under-estimated or over-estimated rotational velocities.

    The ratios \msavre/\TNGvre\ do not show a dependence on intrinsic $C/A$ or $r_{\rm e}$, indicating that poor inferences of \msavre\ are not systematically biased towards small sources or spheroidal ($C/A\gg0.1$) systems. However, the inference fidelity of rotational velocities (\msavre/\TNGvre) shows a mild anti-correlation with the intrinsic rotational support measure, \TNGvre/\TNGsigma\ (Spearman rank coefficient of $\rho=-0.19$), such that the best-fit \msavre\ are over-predicted for low intrinsic \TNGvre/\TNGsigma\ galaxies (both blue and green subsets).
    
    Turning to the inference of $\sigma$, we find that the best-fit \msasigma\ values are on average equal to the intrinsic \TNGsigma, with 16th and 84th percentiles of the ratio \msasigma/\TNGsigma\ being 0.63 and 1.57, respectively. These statistics correspond to a scatter of a factor of $\sim1.5$, i.e., substantially smaller than the scatter found for the rotational velocities. The inference fidelity shows weak anti-correlations with the stellar mass and intrinsic shortest-to-longest axis ratios $C/A$, and no dependence on the effective radius. There is a strong correlation with the intrinsic \TNGvre/\TNGsigma\ (Spearman rank coefficient of $\rho=0.15$), which is in the opposite direction of the anti-correlation found for the rotational velocity. We discuss the possible causes for these trends in Sec.~\ref{sec:discussion}.

\subsection{Redshift evolution of inferred and intrinsic kinematics}
    \label{subsec:kin_vs_z}

    Observational studies of ionised gas find a mild increase of disc turbulence ($\sigma$) in star-forming galaxies with redshift, at least to $z\sim3$ \citep{Uebler_2019}. However, the evolution at earlier times is still unclear, as a wide range of velocity dispersions and trends have been reported in the literature \citep[e.g.][]{deGraaff_2024,Danhaive_2025a,Wisnioski_2025}. Although this may be in part driven by different sample selections, it also calls into question whether the evolution of ionised gas kinematic properties (such as $\sigma$) with redshift could be driven by measurement or modelling uncertainties. 
    
\begin{figure*}
        \centering
        \includegraphics[width=0.85\textwidth]{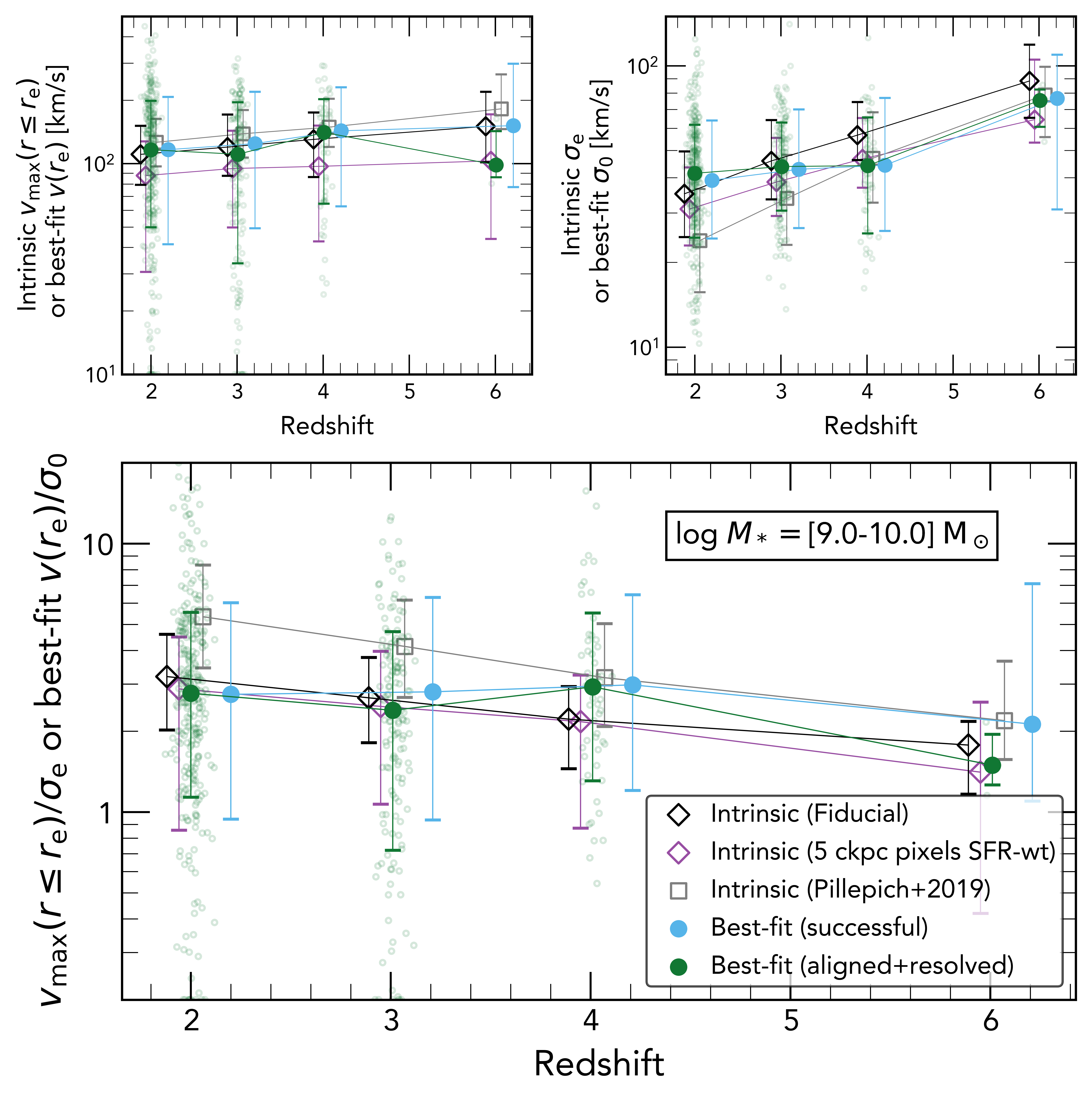}
        \caption{Redshift evolution of the intrinsic and best-fit kinematics for TNG50 galaxies with $M_*=[10^9-10^{10}\,\Msun]$ of the mock NIRSpec sample. The x-axis values are shifted slightly away from the integer redshifts ($z=2,3,4,6$) for visual clarity. The errorbars correspond to 16th and 84th percentiles at fixed $z$. As in Fig.~\ref{fig:TNGvandsigma_vs_z}, the fiducial (and a variation of) intrinsic measurements are shown in black (and purple; legend in bottom panel). The best-fit kinematics for successful fits (i.e., did not hit the bounds) are shown in blue, and those for aligned and resolved fits are shown in green (see Fig.~\ref{fig:SFRMS}). The empty green circles are the individual values for the best-fit aligned$+$resolved fits. (a) Top left: Intrinsic rotational velocity maximum within $r_{\rm e}$ (\TNGvre) or best-fit rotational velocity $r_{\rm e}$ (\msavre). (b) Top right: Intrinsic velocity dispersion within $r_{\rm e}$ (\TNGsigma) or best-fit velocity dispersion (\msasigma). (c) Bottom: Rotational velocity over velocity dispersion \TNGvre/\TNGsigma\,(intrinsic) and \msavre/\msasigma\,(best-fit). Best-fit kinematics agree well with intrinsic measurements on their mild evolution with redshift, but with wider scatter at fixed $z$. \label{fig:kin_vs_z}}
    \end{figure*}

    We therefore assess the evolution of the rotational velocities ($v$), velocity dispersions ($\sigma$), and the degree of rotational support ($v/\sigma$) as a function of redshift in Fig.~\ref{fig:kin_vs_z}, focusing on a single stellar mass bin ($M_*=10^9-10^{10}\,\Msun$). In particular, we compare the intrinsic values measured as in Sec.~\ref{subsec:TNG50_intrinsic} with the best-fit kinematics obtained via forward modelling and fitting of mock NIRSpec/MSA spectra. We show values for successful fits and for fits of aligned$+$resolved sources (see Sec.~\ref{subsec:inference_fidelity} for descriptions). We also include an alternative intrinsic measurement with 10x poorer spatial resolution and SFR-weighting to highlight the variations in the intrinsic measurements from TNG50 (see Sec.~\ref{subsec:TNG50_intrinsic}).

    There is an increase in turbulent motions of the gas (i.e., $\sigma$) with increasing redshift for all (intrinsic and best-fit) measurements, in agreement with some observational studies of ionised gas \citep{Uebler_2019,Danhaive_2025a}. At $z\geq3$, the purple (SFR-weighted and 5 ckpc bins), rather than the fiducial (unweighted and 0.5 ckpc bins) intrinsic \TNGsigma\ agrees better with the best-fit \msasigma, pointing to possible underestimates of turbulent motions beyond cosmic noon. Yet, the best-fit rotational velocities are compatible with the fiducial intrinsic measurement, albeit with large scatter. Importantly, the scatter for best-fit rotational velocities is much larger than the intrinsic, which is not the case for the dispersions.
        
    Importantly, we find good agreement between the best-fit and intrinsic kinematics for the median evolution of the rotational support across $z\sim2-6$. Overall the evolution across redshift is weak, and only the results of \citet{Pillepich_2019} deviate slightly with an upturn in the evolution in $v/\sigma$ toward $z\sim2-3$. However, the scatter in $v/\sigma$ at fixed redshift is much larger for the best-fit values compared to the intrinsic variation. The top panels of Fig.~\ref{fig:kin_vs_z} suggest that the noisy measurement of $v$ primarily leads to wide scatter in $v/\sigma$, whereas the measurement of $\sigma$ affects the median values (in particular, for $z=4$) in addition to introducing scatter. As a result, the scatter in the ratio $v/\sigma$ can be very large, possibly biasing the prevalence of outliers (i.e., dynamically very cold or hot discs) in the underlying galaxy population, discussed in further detail in Sec.~\ref{sec:discussion}.

\section{Velocity dispersion scaling relations}
    \label{sec:sigma_scaling}

    Motivated by the good agreement between the intrinsic and mock-observed trends of the galaxy kinematics as function of redshift in Sec.~\ref{subsec:kin_vs_z}, we next explore the dependence of gas velocity dispersions on galaxy physical properties to assess whether intrinsic trends remain evident for best-fit velocity dispersions from mock spectra. Theoretical models of disc self-regulation indicate that feedback from star formation, along with radial mass transport can drive gas turbulence in galaxies \citep{Toomre_1964,Krumholz_2018}. Observational studies often use such models to interpret their measurements \citep[e.g.][]{Uebler_2019,Girard_2021}, for instance by comparing the velocity dispersion as a function of star formation rate. We therefore investigate variations in $\sigma$ as a function of the global star formation rates, gas fractions, supermassive black hole masses, and accretion rates for our best-fit and intrinsic kinematic properties.
    
\begin{figure*}
        \centering
        \includegraphics[width=\textwidth]{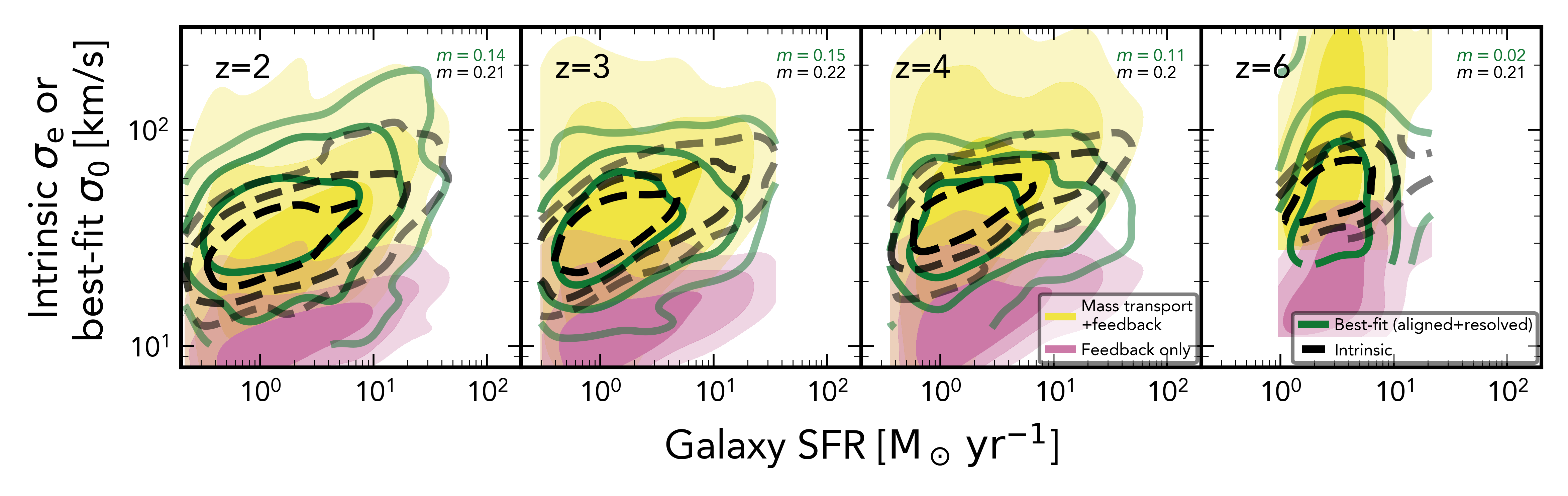}
        \caption{Intrinsic and best-fit velocity dispersions (\TNGsigma\ and \msasigma) vs galaxy star formation rate for TNG50 galaxies with stellar masses of $10^8-10^{11.5}\,\Msun$ at $z=2,3,4,6$. The distributions are shown as contours with increasing transparency, containing values within the 50th, 84th and 97th percentiles. The fiducial intrinsic \TNGsigma\ and best-fit \msasigma\ (aligned$+$resolved) are shown as dashed black and solid green contours, respectively. Theoretical relations of $\sigma$ vs SFR for the mass-transport$+$feedback and feedback-only models of \citet{Krumholz_2018} are shown in yellow and pink, respectively. The contours correspond to a continuum of 50th to 97th percentiles, obtained after using the intrinsic rotation curve velocities of the aligned and resolved sample. The slopes of the linear fits to the intrinsic and best-fit data are indicated in the top-right corner. The distributions of intrinsic and best-fit kinematics are similar, suggesting no strong biases due to star formation on inferred kinematics. \label{fig:sigma_vs_SFR}}
    \end{figure*}

    First, we examine the galaxy velocity dispersion $\sigma$ as a function of the global star formation rate in Fig.~\ref{fig:sigma_vs_SFR}, for the intrinsic (\TNGsigma; dashed black) and best-fit (\msasigma; solid green) \Halpha\ gas velocity dispersions at redshifts $z=2,3,4,6$. The ``mass-transport and feedback" and ``feedback-only" models of \citet{Krumholz_2018} are shown as yellow and pink shaded regions, respectively. These correspond to intrinsic rotation velocities \TNGvre\ ($v_\phi$ in the notation of \citealt{Krumholz_2018}) of the best-fit aligned$+$resolved sample. We make linear fits to the $\sigma-$SFR relation in log-log space, and indicate the resulting slopes in the top right of each panel (colour-coded by intrinsic and best-fit).

    In both the best-fit and intrinsic cases, the gas velocity dispersions increase monotonically with SFR in all redshift bins, and their distributions (solid green, dashed black) are well aligned. Crucially, this implies that the inference fidelity of \msasigma\ does not depend on SFR, and that the overall trend is well recovered from mock observations. The galaxy-to-galaxy scatter at fixed SFR is larger in the case of the best-fit \msasigma, as may be expected from observational and model uncertainties. We note that a larger fraction of galaxies have low best-fit velocity dispersions at fixed SFR as shown by the 84th and 97th percentile contours, especially at $z\geq4$. Although the $\sigma$-SFR relation is marginally flatter (has shallower slopes) for the best-fit \msasigma, both the best-fit and intrinsic distributions agree well with the ``mass-transport+feedback" model of \citet{Krumholz_2018}.

\begin{figure*}
        \centering
        \includegraphics[width=\textwidth]{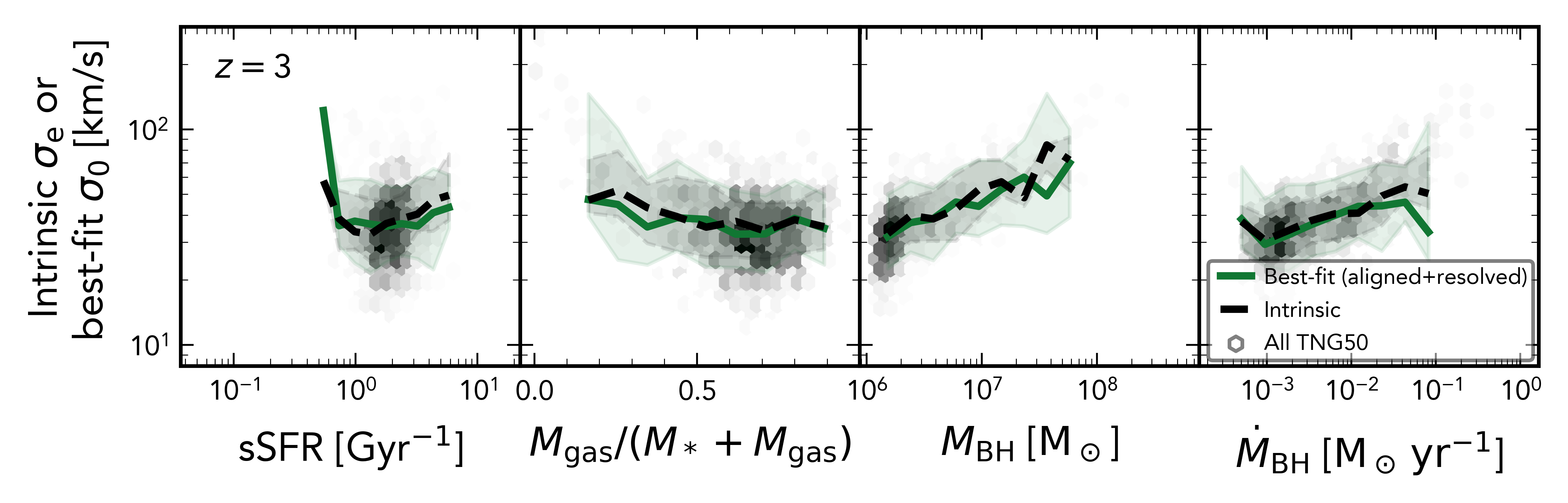}
        \caption{Intrinsic and best-fit velocity dispersions (\TNGsigma\ and \msasigma) vs galaxy specific star formation rate (sSFR), gas fraction ($f_{\rm gas}=M_\mathrm{gas}/(M_*+M_\mathrm{gas})$), supermassive black hole mass ($M_{\rm BH}$), and accretion rate ($\dot{M}_\mathrm{BH}$) for $z=3$ TNG50 galaxies. Running medians for the fiducial intrinsic \TNGsigma\ and best-fit \msasigma\ (aligned$+$resolved) are shown as dashed black and solid green lines, respectively. Shaded regions correspond to area within the 16th and 84th percentiles at each x-axis bin. Slopes to linear fits in log-log space are indicated in the top-right of each panel, colour-coded by intrinsic and best-fit. The hexbins show the intrinsic \TNGsigma\ values for all TNG50 galaxies with $M_*>10^8\,\Msun$, colour-coded in grey by the number of galaxies in each hexbin. Intrinsic and best-fit kinematics show, on average, similar relations with various global galaxy properties. \label{fig:sigma_vs_galprop}}
    \end{figure*}

    Next, we consider only galaxies at $z=3$ and correlate $\sigma$ to galaxy properties. In Fig.~\ref{fig:sigma_vs_galprop}, we show the best-fit \msasigma\ and intrinsic \TNGsigma\ as a function of specific star formation rate (sSFR), gas fraction ($f_{\rm gas}=M_\mathrm{gas}/(M_*+M_\mathrm{gas})$), SMBH mass ($M_{\rm BH}$), and SMBH accretion rate ($\dot{M}_\mathrm{BH}$). Overall, the distributions of best-fit and intrinsic velocity dispersions align well with each other across all four galaxy physical properties. We find no dependence on sSFR for the best-fit \msasigma\ and intrinsic \TNGsigma. The velocity dispersions are anti-correlated with the gas fractions for $f_{\rm gas}\lesssim0.4$ but flatten to $\sim30\,\kms$ for larger values. Although the increase in $\sigma$ for low gas fractions follows an increase in stellar mass, secondary effects due to larger mass transport may act on the velocity fields \citep{Krumholz_2018}. We also find that the velocity dispersions increase marginally with SMBH mass and accretion rate, although these properties also correlate with stellar mass and therefore may not be a causal effect.
    
    In summary, we find that global scaling relations of velocity dispersions with various galaxy properties are recovered well through fits to mock spectra. Despite a scatter of $\sim$1.5 in the inference fidelity of $\sigma$ (for $z=3$; see Sec.~\ref{subsec:inference_fidelity}), the intrinsic and best-fit values have statistically similar distributions across star formation rates, gas fractions, and supermassive black hole properties. Our results suggest that observationally derived kinematics can be reliable tools for constraining the effects of feedback and gravity on driving turbulence in early galaxies.

\section{Discussion}
    \label{sec:discussion}

    In Sec.~\ref{sec:inference_results}, we showed that kinematic parameters recovered from mock JWST NIRSpec spectra are on average consistent with the intrinsic simulation values of TNG50 star-forming galaxies. However, the best-fit $v$ and $\sigma$ exhibit wide scatter about the average (see Fig.~\ref{fig:kin_vs_z}). In this Section, we discuss the impact of these results on quantifying $v/\sigma$ beyond cosmic noon, as well as the efficacy of simple kinematic models in describing intrinsically complex galaxy kinematics. 

    \subsection{Rotational support at cosmic noon}
        \label{sec:discussion_vsigma}

    Using the mock NIRSpec/MSA spectra and thin disc kinematic modelling, we evaluated rotational velocities and dispersions for hundreds of star-forming galaxies at cosmic noon. This approach allows us to quantify for the first time, using mock JWST observations, the evolution of rotational support ($v/\sigma$) of simulated TNG50 galaxies from $z=2-6$. Crucially, we test whether the inferred \msavre/\msasigma\ reflects the intrinsic ``diskiness" of star-forming gas in high-redshift galaxies as measured directly from the simulation. 
    
\begin{figure}
        \centering
        \includegraphics[width=\columnwidth]{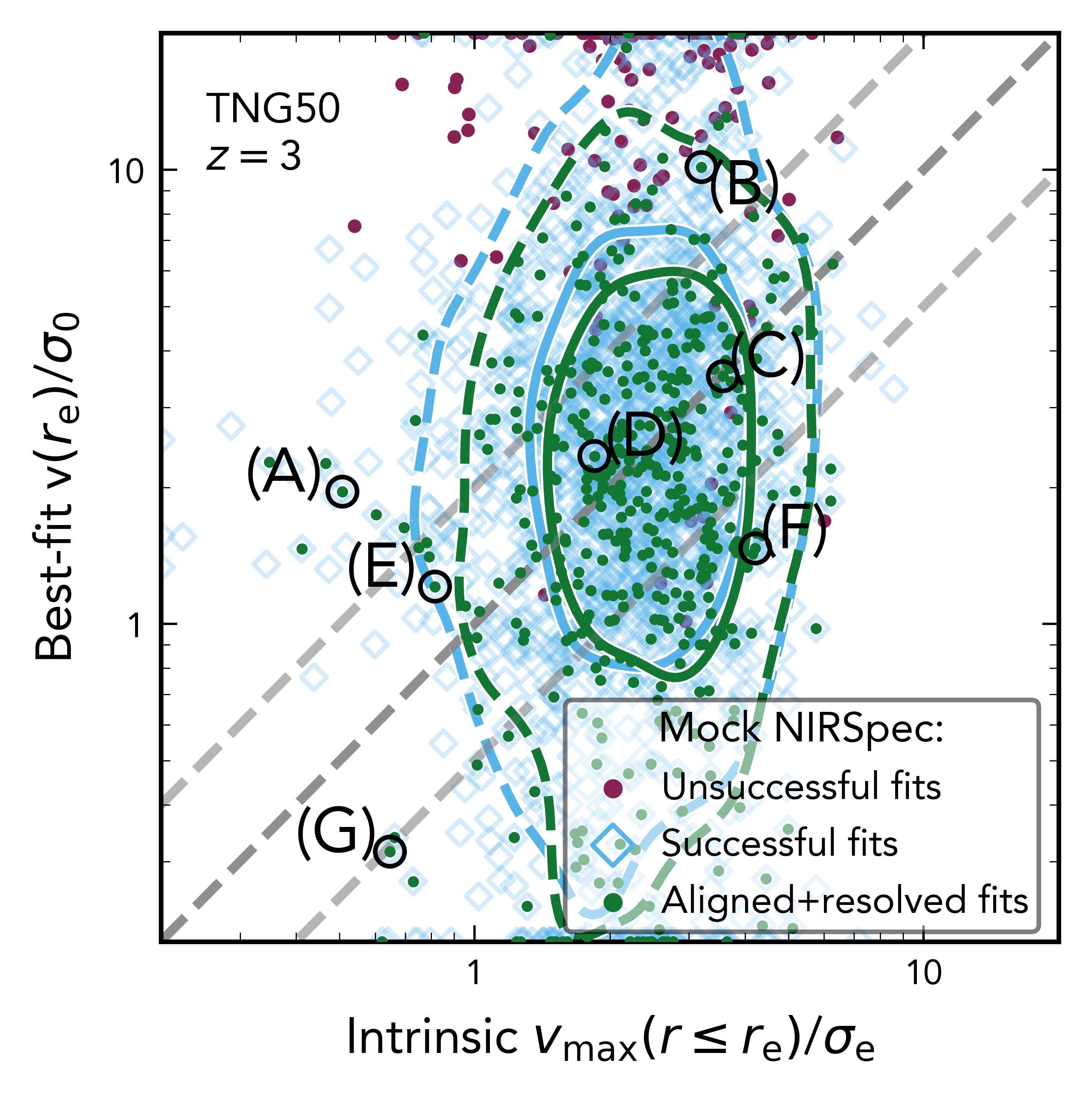}
        \caption{Best-fit $v(r_{\rm e})/\sigma_0$ vs intrinsic $v_{\rm max}(r\leq r_{\rm e})/\sigma_e$ for TNG50 galaxies at $z=3$ with stellar masses of $10^8-10^{11.5}\,\Msun$. The different lines and colours correspond to subsets as in Fig.~\ref{fig:bestfitvsintrinsic} (see legend). Seven example fits to the mock imaging and spectra are labelled as (A) to (G), corresponding to Fig.~\ref{fig:collage_vbysigma}.}
        \label{fig:bestfitvsintrinsicvbysigma}
    \end{figure}

    In Fig.~\ref{fig:bestfitvsintrinsicvbysigma}, we compare the best-fit $v(r_{\rm e})/\sigma_0$ to the intrinsic $v_{\rm max}(r\leq r_{\rm e})/\sigma_{\rm e}$ from TNG50 for $z=3$. Best-fit $v(r_{\rm e})/\sigma_0$ values larger than the visualised limit are placed at the top of the plot. The bulk of the galaxies lie in the region spanning $v/\sigma=[1,3]$, with a large vertical scatter. The contours also cover ranges above and below the 1:2 and 1:1/2 lines, implying that $v/\sigma$ can be off by a factor of more than 2 in up to a third of sources. We find only slightly larger scatter for the successful fits compared to the fits for aligned$+$resolved sources. The large scatter of points above and below the 1:1 line, especially for systems with intrinsic \TNGvre/\TNGsigma$\lesssim3$ highlight the possibility of misclassifying individual systems as highly rotationally supported. In contrast, there are also systems which have large intrinsic \TNGvre/\TNGsigma, but have underestimated \msavre/\msasigma.

    Our findings indicate that for $z\geq2$ TNG50 galaxies, it is possible to constrain $v/\sigma$ with a thin disc model only within a factor of $\sim2-3$ on average. We emphasise that this is the case even for sources that are well-aligned and resolved in single slit MSA spectra with high signal-to-noise ratios. We have further tested the effect of observational uncertainties in Appendix~\ref{sec:appendix_kin_model} (see Fig.~\ref{fig:noise_comparison}), but find that improving the $(S/N)$ by a factor two does not improve the recovered $v/\sigma$ values. We conclude that determining the ``diskiness" of galaxies is extremely challenging for individual objects, at least when using simple thin disc models, and that ensembles of galaxies are crucial to draw meaningful conclusions. However, we highlight that the dynamic range in $v/\sigma$ in TNG50 at high redshifts is small, and it is therefore difficult to assess whether these conclusions also hold for galaxies with intrinsically strong rotational support (i.e. $v/\sigma>5$).
    
    Although we do not analyse $z>6$ galaxies in this work, we expect that observations of cosmic dawn galaxies will suffer from similar biases as those discussed above. We therefore caution against over-interpreting the rotational support measured for individual systems \citep[as also discussed in][]{deGraaff_2024}, although population trends could still be well recovered. To improve upon the large scatter found in our analysis, precisely quantifying the level of rotational support in early galaxies from observations will likely require moving beyond 2D spectroscopy to 3D IFU \citep{Perna_2023,Parlanti_2024, Arribas_2024} or slit-stepping strategies \citep{Barisic_2025,Morrison_2025}. However, this may remain challenging even with IFU data at hand if galaxies have inherently complex kinematics, as demonstrated recently by \citet{Phillips_2025}.
    
    \begin{figure*}
        \centering
        \includegraphics[width=\textwidth]{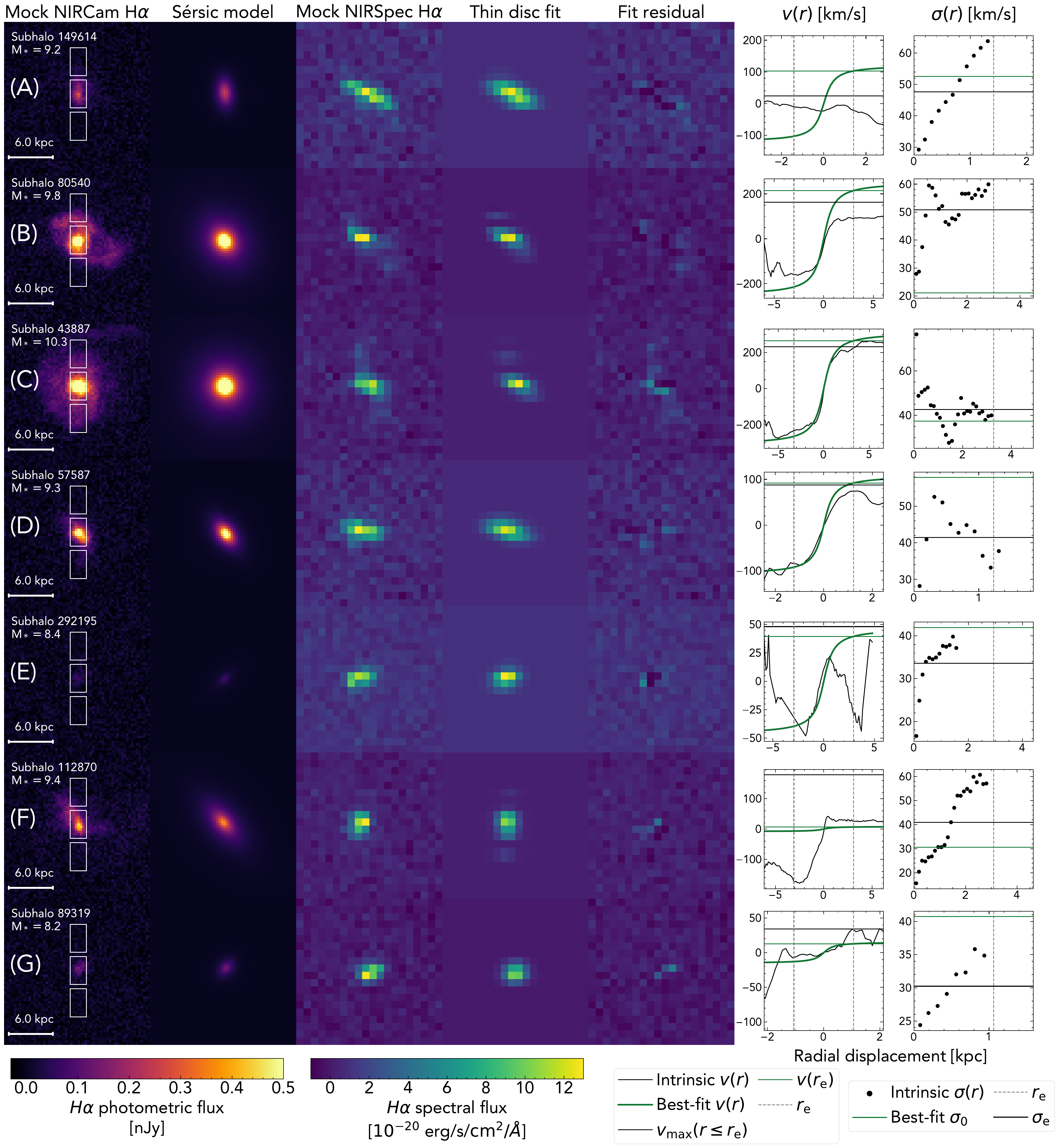}
        \caption{Collage of fit \Halpha\ photometry and kinematics for $z=3$ TNG50 galaxies mapped from their positions in the best-fit \msavre/\msasigma-\TNGvre/\TNGsigma\ space. The seven columns correspond to NIRCam mock \Halpha\ images, \Sersic\ fit model to the mock photometry, NIRSpec mock \Halpha\ spectra, Thin disc kinematic model fit to the spectra, kinematic model fit residuals, intrinsic and best-fit rotation velocity profiles, and intrinsic and best-fit velocity dispersion profiles, respectively. The stellar masses and scales are indicated in the first column of each TNG50 galaxy. The NIRSpec mock slit is placed vertically in each case as indicated by the white rectangles. The galaxies shown here are those labelled as (A)-(G) in Fig.~\ref{fig:bestfitvsintrinsicvbysigma}, placed in descending order of estimation, i.e., overestimated (top; A and B) to well-recovered (middle; C, D, and E) to underestimated (bottom; F and G). The best-fit and intrinsic kinematics may not always be fully informative despite being recovered accurately.}
        \label{fig:collage_vbysigma}
    \end{figure*}

    \subsection{Efficacy of thin disc models}
    
    In this work, we model galaxy kinematics to mock spectra by assuming a simple parametric model of a rotating disc that is infinitely thin. However, we find broad distributions of shortest-to-longest axis ratios for TNG50 galaxies in our sample (see Fig.~\ref{fig:bestfitbyintrinsic_vs_galprop} and \ref{fig:bestfitbyintrinsic_vs_galprop_altz}). To investigate the validity of our kinematic modelling assumptions, we compare intrinsic and best-fit velocity profiles in Fig.~\ref{fig:collage_vbysigma}. We show rotational velocity and velocity dispersion profiles for seven galaxies at $z=3$ that span the $v/\sigma$ plane of Fig~\ref{fig:bestfitvsintrinsicvbysigma}, along with their mock NIRCam images, \Sersic\ models, mock NIRSpec spectra, best-fit model spectra, and fit residuals. The galaxies are chosen to represent overestimated (top two rows), well-recovered (middle three rows), and underestimated (bottom two rows) $v/\sigma$.
    
    From the middle three rows (C, D, E), i.e., where the kinematic properties are recovered well, we find that the thin disc model is sufficient to describe the rotational velocity maximum (\TNGvre) and average velocity dispersions (\TNGsigma). However, the intrinsic velocity dispersion profile is complex, and the intrinsic \TNGsigma\ itself fails to capture the variations due to localised, radial differences. Given this limitation, the best-fit \msasigma\ can optimally only be as informative as the intrinsic \TNGsigma, and may be biased above or below this value. Similarly, the intrinsic rotational velocity profile and hence the mock spectra can deviate from the arctangent profiles used in our parametrisation \citep{Courteau_1997} (see, for e.g. row E in Fig.~\ref{fig:collage_vbysigma}). We caution that, to measure these intrinsic kinematics, we used a projection that assumes the morphological and kinematic axes are equivalent. Although this assumption is practical for galaxies with no well-defined kinematic axes (as is the case for low intrinsic $v/\sigma$ galaxies), their morphologies may still not reflect the orientation of the dominant velocity components. Together, these deviations underscore the limitations of using single measurements to interpret complex galaxy kinematics, even for high-fidelity spectral fits or the intrinsic quantities measured from simulations.
    
    In cases of over/under-estimated velocities, we also find asymmetries in the intrinsic rotational velocity and dispersion profiles (see for e.g. row B in Fig.~\ref{fig:collage_vbysigma}). Non-circular motions, inflows and outflows can cause these features, such that even intrinsic \TNGvre\ measurements of the star-forming gas do not trace the gravitational potential of galaxies. Cosmological simulations such as TNG50 predict faster galactic outflows at larger stellar masses and at higher redshift \citep{Nelson_2019b}. Observational evidence indeed also points to a high prevalence of outflows of gas from star-forming galaxies even at $z>2$ \citep{Davies_2019,Weldon_2024,Carniani_2024,Del_Pino_2026}, which can thus contaminate rotational measures of the ionised gas. Moreover, gravitational perturbations can induce non-circular motions, and thereby deviations from the simple parametrisation used in our forward modelling. Such perturbations are often linked to clumpy substructures in simulations \citep[e.g.][]{Mandelker_2014}; although commonly seen in high-redshift galaxies \citep[e.g.][]{Claeyssens_2025,delaVega_2025}, their effect on the observed galaxy-scale kinematics appears modest \citep[e.g.][]{Genzel_2011,Rizzo_2024}. 
    
    In addition to the physical complexity of intrinsic kinematic profiles that cannot be captured by our modelling, observational effects may also play an important role. For instance, if the galaxy is either very compact compared to the PSF FWHM, or very extended compared to the NIRSpec slit, $v$ may be incorrectly estimated due to not resolving the true rotation curve maximum or poorly estimating the turnover radius $r_{\rm t}$ (see bottom two panels in Fig.~\ref{fig:collage_vbysigma}). Furthermore, because we fit our mock spectra assuming an infinitely thin disc (i.e. $C/A\sim0$), we use the observed axis ratio to correct for inclination in the velocity profiles. In Fig.~\ref{fig:bestfitbyintrinsic_vs_galprop}, we showed that the best-fit \msasigma\ are underestimated for larger intrinsic shortest-to-longest axis ratios $C/A$. We also found that for galaxies with intrinsic \TNGvre/\TNGsigma$<4$, rotational velocities are overestimated and velocity dispersions are underestimated. For such systems with large intrinsic $C/A$, the inclination correction to the velocity profile as derived from the observed axis ratio $q$ causes $v$ to be overestimated. To compensate, the fit typically underestimates the $\sigma$ of the system. For example, a spheroidal system would be modelled as a face-on galaxy in kinematic fits assuming infinitely thin discs. Our findings highlight the merit of linking galaxy morphologies to kinematic properties at cosmic noon and beyond \citep{Price_2016,Espejo_Salcedo_2025,Danhaive_2025b}, while connecting such intrinsic galaxy properties to biases in dynamical inference.

    Finally, the chosen weighting scheme also affects the efficacy of estimating $v$ and $\sigma$. In our weighted $\chi^2$ method Sec.~\ref{sssec:kin_model}), we up-weight the low SNR rows to pick up rotation or dispersion features from fainter parts of the spectrum. However, these features are sensitive to spurious features in the mock spectra that deviate from disc-like emission profiles. The weighting scheme thus causes our single-slit spectral fits to be biased towards higher $\sigma$ for galaxies with lower stellar masses, which are typically fainter. We performed tests with other weighting schemes (such as $\sqrt{SNR}$-weighted or unweighted) and found no statistical differences, although individual fits behaved differently (see Fig.~\ref{fig:SNRwt_comparison}).
    
    To identify causes for unsuccessful fits to the mock spectra (magenta points, as labelled in Fig.~\ref{fig:SFRMS}), we selected galaxies for which the kinematic fit \msasigma\ hit the lower bound of $10\,\kms$, which comprise the majority of this subset. We found that nearly all of these mock sources have poorly constrained morphologies from \pysersic\ or have slits misaligned from the rotation axis of the galaxy (we recall that the slit is placed North-South for all sources). For $\sim1$\% of sources whose morphological properties were well-constrained (i.e., narrow percentile widths for $q$, PA, \& $r_{\rm e}$) and the best-fit kinematics have low $\chi^2$, the intrinsic \TNGsigma\ falls below the detectable threshold, causing the kinematic model to indicate a \msasigma\ of $\sim10\,\kms$. We explored the effect of using a smaller value for the \msasigma\ lower bound but found that our modelling could not reliably recover dispersions below $10\,\kms$, implying that either the spectral resolution ($\sim30-50\,\kms$) is too low or the thin disc assumption is insufficient to model galaxies with intrinsically low \TNGsigma. 

    \subsection{Implications for high-redshift gas kinematics}
    
    In Sec.~\ref{subsec:kin_vs_z}, we showed that the marginal increase of the intrinsic $v/\sigma$ (of $M_*=[10^9-10^{10}\,\Msun]$ galaxies) with decreasing redshift is recovered by the kinematic modelling. However, there is a large uncertainty for individual sources at $z=3$ (Sec.~\ref{sec:discussion_vsigma}), and we therefore also explore the ratio of the best-fit to intrinsic kinematics as a function of galaxy properties for $z=2,4,6$ in Fig.~\ref{fig:bestfitbyintrinsic_vs_galprop_altz}. We find that the best-fit rotational velocities (velocity dispersions) are overestimated (underestimated) for galaxies with low intrinsic \TNGvre/\TNGsigma, more commonly found at higher redshift (see bottom right panels). Furthermore, the \msasigma\ is underestimated at higher redshift, owing to galaxies typically having larger intrinsic shortest-to-longest axis ratios $C/A$ (see bottom left panels of \msasigma/\TNGsigma) \citep{Pillepich_2019}. Only for the lowest stellar masses ($<10^{8.5}\,\Msun$) are both $v$ and $\sigma$ slightly overestimated on average (top left panels).

    Dispersion-supported systems at higher redshift are hence more likely to be misclassified as rotationally-supported ($v/\sigma\gtrsim2$). We reiterate that the intrinsic global \TNGsigma\ may not describe well the radial changes in $\sigma$ across the galaxy, and thereby neither can the constant best-fit \msasigma. This observational bias towards lower velocity dispersions and higher rotational velocities may arise from a combination of factors, including (i) the assumption of an infinitely thin disc being incompatible with large intrinsic $C/A$; (ii) declining spatial resolution; (iii) intrinsic $v(r)$ and $\sigma(r)$ deviating from an arctangent profile and a constant value. These different effects are likely increasingly important at higher redshifts due to inflows and clumpy star formation (see Fig.~\ref{fig:collage_vbysigma}, and \citealt{Weldon_2024,Carniani_2024,delaVega_2025}). Accounting for these observational and modelling biases is therefore critical, and model testing with hydrodynamical simulations (as in this work) provides an important path forward to improving observational constraints.
    
\section{Conclusions}
    \label{sec:conclusions}
    
    We used the TNG50 simulation \citep{Nelson_2019b,Pillepich_2019} to construct mock observations of 5969 star-forming galaxies at $z=2-6$ with stellar masses of $10^8-10^{11.5}\,\Msun$. We created mock JWST NIRCam images and NIRSpec/MSA spectra of the simulated \Halpha-emitting gas using \msafit\ \citep{deGraaff_2024} --- a forward modelling and fitting tool for the NIRSpec MSA. We then evaluated morphological and kinematic properties from these mock observations by fitting them as thin, rotating discs. We thereby obtained rotational velocities ($v$) and velocity dispersions ($\sigma$) for hundreds to thousands of galaxies at each redshift.
    
    To test the efficacy of fitting simplistic parametric models (i.e. a thin disc) to galaxies with intrinsically complex kinematic structures, we compared the best-fit kinematic properties $v$ and $\sigma$ with those measured directly from the simulation. We chose, as the default, intrinsic measures conceptually similar to observational inferences, e.g., measured within the half-light radius of the ionised gas emission. However, we stress that different intrinsic measurement choices of $v$ and $\sigma$ of the simulated star-forming gas in TNG50 galaxies can vary by up to a factor of $\lesssim$2 at fixed redshift, depending on choices of aperture, spatial binning and SFR-weighting (Fig.~\ref{fig:TNGvandsigma_vs_z}). 
    
    We summarise our main findings on the comparison of best-fit and intrinsic kinematic properties as follows:
    
    \begin{itemize}
        \item $v$ and $\sigma$ recovered from fits to mock NIRSpec/MSA spectra correspond well to intrinsic measurements, with median ratios of $\sim$1 for best-fit to intrinsic kinematic properties. The scatters in recovery accuracy are e.g., $\sim$2 for $v$ and $\sim$1.5 for $\sigma$ at $z=3$. The scatter for $\sigma$ is symmetric, while $v$ is skewed towards lower values.
            
        \item The recovery of the best-fit parameters is robust for intrinsically disk-like and smooth systems. Additionally, the kinematic modelling works well even for faint sources or complex systems with clumpy light distributions due to our weighted $\chi^2$ fitting (see Fig.~\ref{fig:collage}), allowing us to reproduce the kinematic complexity of a variety of systems.
            
        \item However, best-fit rotational velocities (velocity dispersions) are more likely to be overestimated (underestimated) for galaxies with low intrinsic $v/\sigma$, and this trend is stronger at higher redshift (Fig.~\ref{fig:bestfitbyintrinsic_vs_galprop}). Furthermore, best-fit velocity dispersions are underestimated for larger intrinsic axis ratios ($C/A$). These ratios, i.e., the accuracy of recovery of kinematic properties do not show any dependence on galaxy size ($r_{\rm e}$) and are only weakly dependent on stellar mass.
    
        \item There is intrinsically weak evolution in the median $\sigma$ and $v/\sigma$ across $z=2-6$ in the TNG50 simulation at fixed stellar mass. Our best-fit values obtained from the mock spectra accurately recover this evolution on average. Only at $z=4$ do we find that the median best-fit values for $v/\sigma$ are larger than those of the intrinsic by a factor of $\lesssim2$, primarily due to the underestimation of $\sigma$ for galaxies not resembling thin, rotating discs (Fig.~\ref{fig:kin_vs_z}).
            
        \item At all redshifts, $v/\sigma$ is recovered only with an accuracy of a factor of $2-3$ for individual galaxies (Fig.~\ref{fig:bestfitvsintrinsicvbysigma}), especially for systems with intrinsically low $v/\sigma\lesssim3$. The primary causes of this large scatter are deviations from disc-like geometry (increasingly important at higher redshifts), and radial variations in $\sigma$. 
    
        \item Both the best-fit and intrinsic $\sigma$ show the same dependence on galaxy SFR, and align closely with the ``mass-transport$+$feedback" model of \citet{Krumholz_2018}. Importantly, this indicates that physical inferences of feedback from kinematic modelling are qualitatively reliable. Additionally, intrinsic and best-fit dispersions have similar distributions across galaxy properties such as gas fractions, supermassive black hole masses and accretion rates.
    \end{itemize}
    
    We defer a similar comparative analysis for parameters derived from the kinematic models, specifically the dynamical mass and mass budget, to future work. Nonetheless, for ordered, disc-like galaxies, we expect the uncertainties in derived dynamical masses to be low, given the well-recovered kinematics for these systems.
    
    An important caveat in our work is the underlying assumption that TNG50 galaxies are an appropriate laboratory for testing the kinematic modelling of high-redshift galaxies. At $z>3$, there are only few galaxies in the TNG50 simulation with strong rotational support ($v/\sigma\gtrsim5$), limiting the extent to which we can draw conclusions regarding the fidelity of the modelling. Moreover, choices in the measurement of intrinsic parameters (e.g. the aperture size and projection used) and mock procedure may affect some of our conclusions. For instance, we do not forward model the \Halpha\ emission self-consistently, and ignore the effects of circumgalactic medium gas and strong outflows. These aspects could be improved upon with better post-processing \citep{Hirschmann_2017,Hirschmann_2023,Choustikov_2025,Yang_2025}. Recent and upcoming numerical studies that simulate the multiphase ISM (rather than an effective equation of state as in TNG50) will help overcome these limitations  \citep[e.g.][]{Katz_2022,Katz_2024,Bhagwat_2024,Kannan_2025}.

    Nevertheless, our work crucially presents a general framework that allows for detailed testing of the inference of kinematic properties from NIRSpec MSA spectroscopy, providing a path forward to improving observational constraints and end-to-end comparisons of cosmological simulations. Although we have shown that rotational velocities ($v$) and velocity dispersions ($\sigma$) are recovered well on average when using a simple thin disc model, there can be large uncertainties in the best-fit kinematic properties for individual objects due to the inherent complexity of high-redshift galaxies. To improve upon these model uncertainties will require implementing models that better capture the wide range in 3D morphologies and clumpiness of high-redshift galaxies, informed by empirical constraints on galaxy morphologies \citep[e.g.][]{Pandya_2024,Claeyssens_2025} and tested with hydrodynamical simulations of high-redshift galaxies. These measurements may further benefit from spatially resolved spectroscopy and improved modelling strategies. Specifically, the use of complex 3D kinematic modelling \citep[e.g.][]{Lee_2025a} in combination with IFU \citep[e.g.][]{Arribas_2024,Del_Pino_2026} and MSA slit-stepping programs \citep{Barisic_2025,Morrison_2025} will help exploit the immense capabilities of JWST to probe the kinematic structures of large samples of typical ($M_*\sim10^{7-10}\,\Msun$) star-forming galaxies into the first billion years.

\section*{Acknowledgements}

RA and AP acknowledge funding from the 
European Union (ERC, COSMIC-KEY, 101087822, PI: Pillepich). AdG acknowledges support from a Clay Fellowship awarded by the Smithsonian Astrophysical Observatory. We thank Marijn Franx and Hannah {\" U}bler for helpful comments. Computations were performed on the HPC system Vera at the Max Planck Computing and Data Facility.

\section*{Data Availability}

The data products produced for this study (inferences from mock JWST observations of TNG50 galaxies) are available on request from the corresponding author. All data pertaining to the IllustrisTNG project, including TNG50, are openly available \citep{Nelson_2019a} and can be retrieved from the IllustrisTNG website at \href{www.tng-project.org/data}{www.tng-project.org/data}.

\bibliographystyle{mnras}
\bibliography{main.bib}
\appendix
\section{Kinematic modelling variations}
\label{sec:appendix_kin_model}

\begin{figure}
    \centering
    \includegraphics[width=0.45\textwidth]{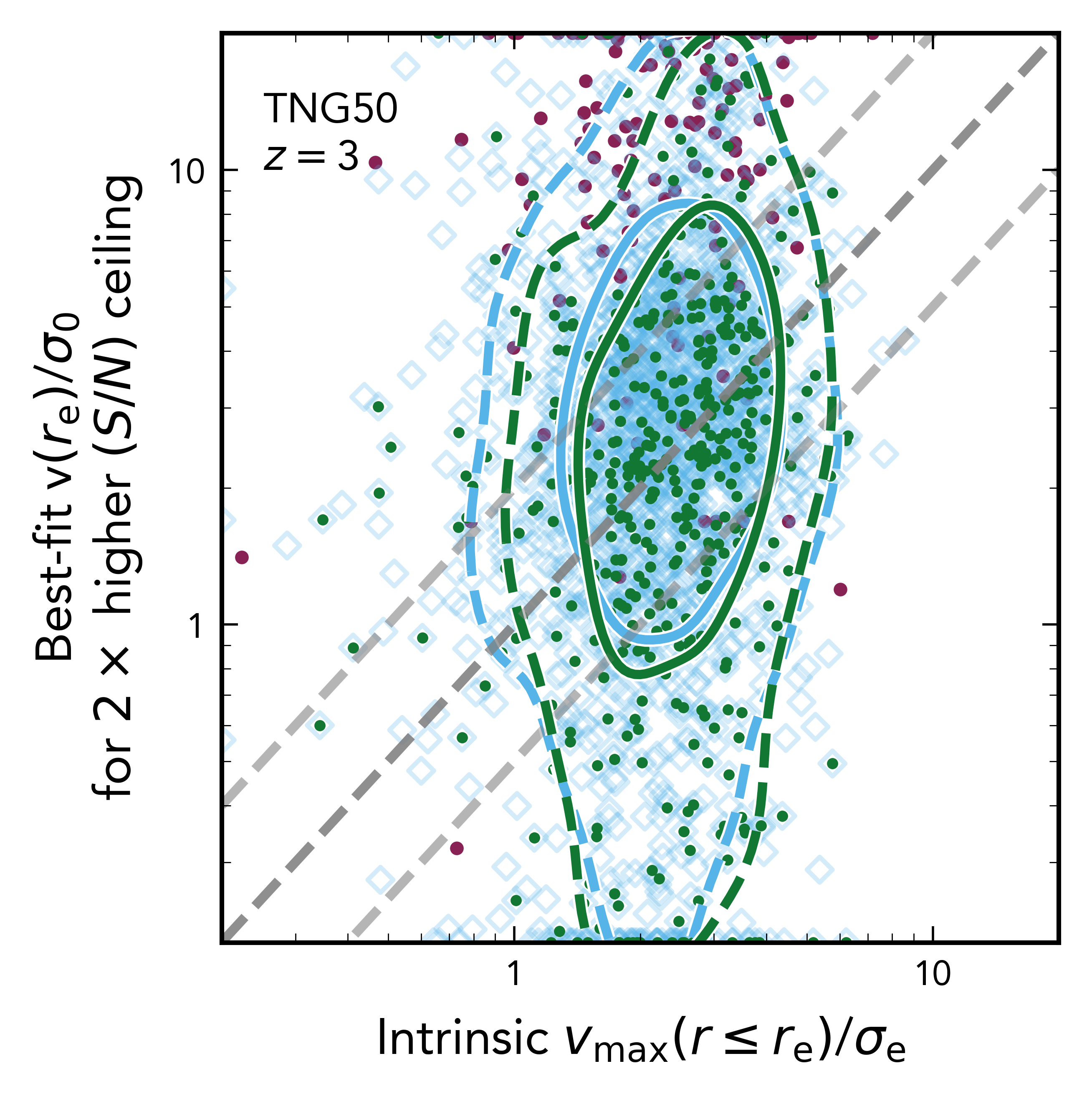}
    \caption{Best-fit $v(r_{\rm e})/\sigma_0$ vs intrinsic $v_{\rm max}(r\leq r_{\rm e})/\sigma_e$ for $z=3$ TNG50 galaxies, but for mock NIRSpec/MSA spectra with twice the fiducial $(S/N)$ ceiling, where the fiducial corresponds to a minimum $\sigma_{\rm noise}$ of $4\times10^{-21}\,\mathrm{erg/s/cm^2/}$\AA\ (see Sec.~\ref{sssec:nirspec}). The lines and annotations are as in Fig.~\ref{fig:bestfitvsintrinsicvbysigma}.}
    \label{fig:noise_comparison}
\end{figure}

In this Section, we test if the best-fit kinematic properties change for different variations of the mocking and fitting procedure (as described in Sec.~\ref{sssec:nirspec} and \ref{sssec:kin_model}). First, we make mock JWST NIRSpec/MSA spectra of $z=3$ TNG50 galaxies with \Halpha\ $(S/N)>10$ in the mock NIRCam imaging. However, we now use twice the fiducial $(S/N)$ ceiling, i.e., a factor of $2$ lower Gaussian noise $\sigma_{\rm noise}$ (see Sec.~\ref{sssec:nirspec}). We then fit these spectra with thin-disc models to obtain rotational velocities ($v$) and velocity dispersions ($\sigma$). In Fig.~\ref{fig:noise_comparison}, we compare intrinsic measurements to best-fit kinematic properties obtained with these mock spectra. We find no statistical differences in the recovery of \msavre\ and \msasigma\ in comparison to the intrinsic values, despite a two-fold increase in the $(S/N)$ of the mock spectra.

\begin{figure}
    \centering
    \includegraphics[width=0.5\textwidth]{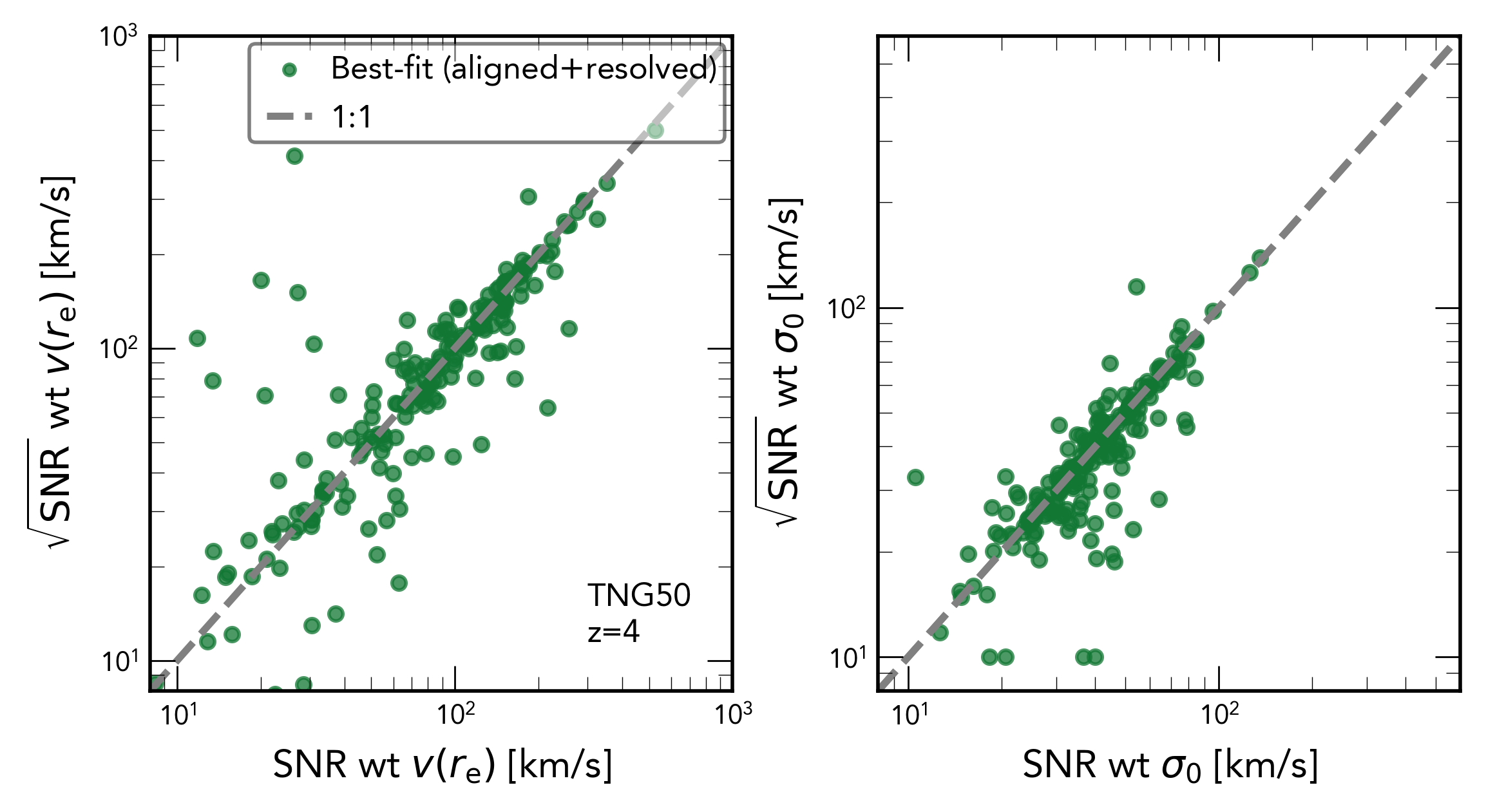}
    \caption{Comparison of best-fit (aligned$+$resolved) kinematics obtained using different $(S/N)$ weighting schemes (see Sec.~\ref{sssec:kin_model}) for $z=3$ TNG50 galaxies with stellar masses of $10^8-10^{11.5}\,\Msun$. The x-axis and y-axis show the kinematics for inverse SNR weighting (i.e., the fiducial choice $w_y=[(S/N)(y)]^{-1}$) and inverse square root SNR weighting (i.e., $w_y=[(S/N)(y)]^{-0.5}$).}
    \label{fig:SNRwt_comparison}
\end{figure}

Next, we fit the fiducial mock spectra using the weighted-$\chi^2$ method (described in Sec.~\ref{sssec:kin_model}), but now changing the weighting scheme to be inversely proportional to the square root of the $(S/N)$. In Fig.~\ref{fig:SNRwt_comparison}, we compare the best-fit \msavre\ and \msasigma\ for the kinematic modelling with the fiducial and modified weighting schemes. We find no substantial differences in the best-fit kinematics across weighting choices, although there may be discrepancies at low \msavre\ (see the left panel).

\onecolumn

\section{Inference fidelity for alternate redshifts}
\label{sec:appendix_altz}

\begin{figure*}
    \centering
     \includegraphics[width=0.32\textwidth]{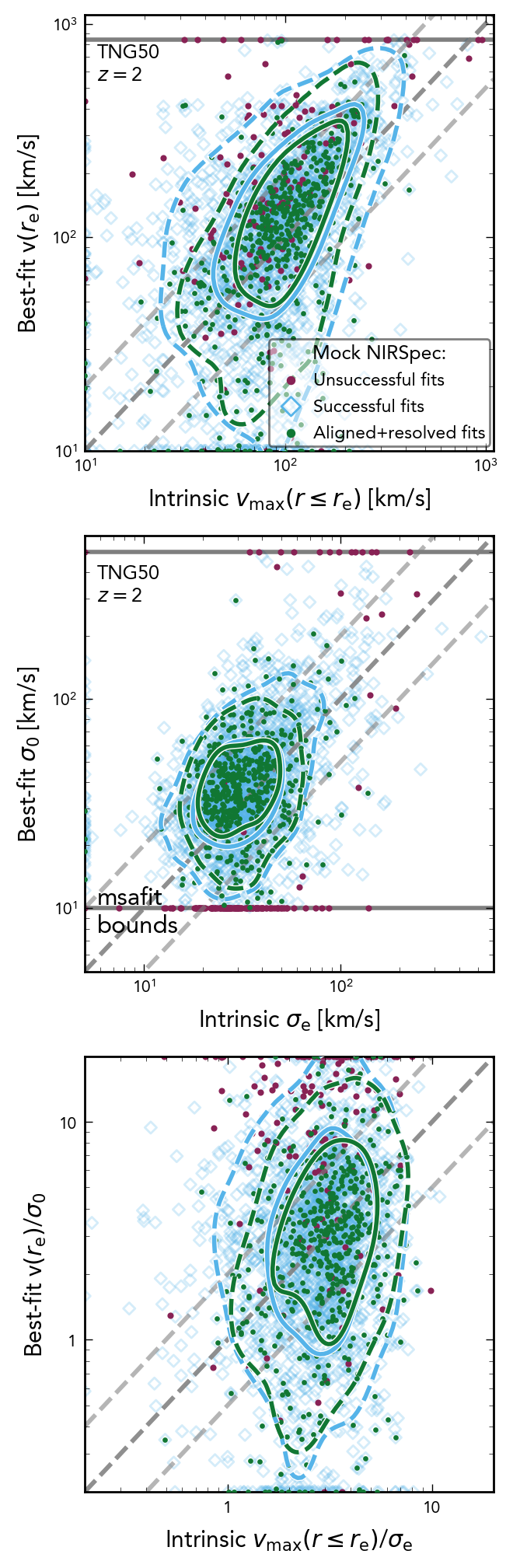}
    \includegraphics[width=0.32\textwidth]{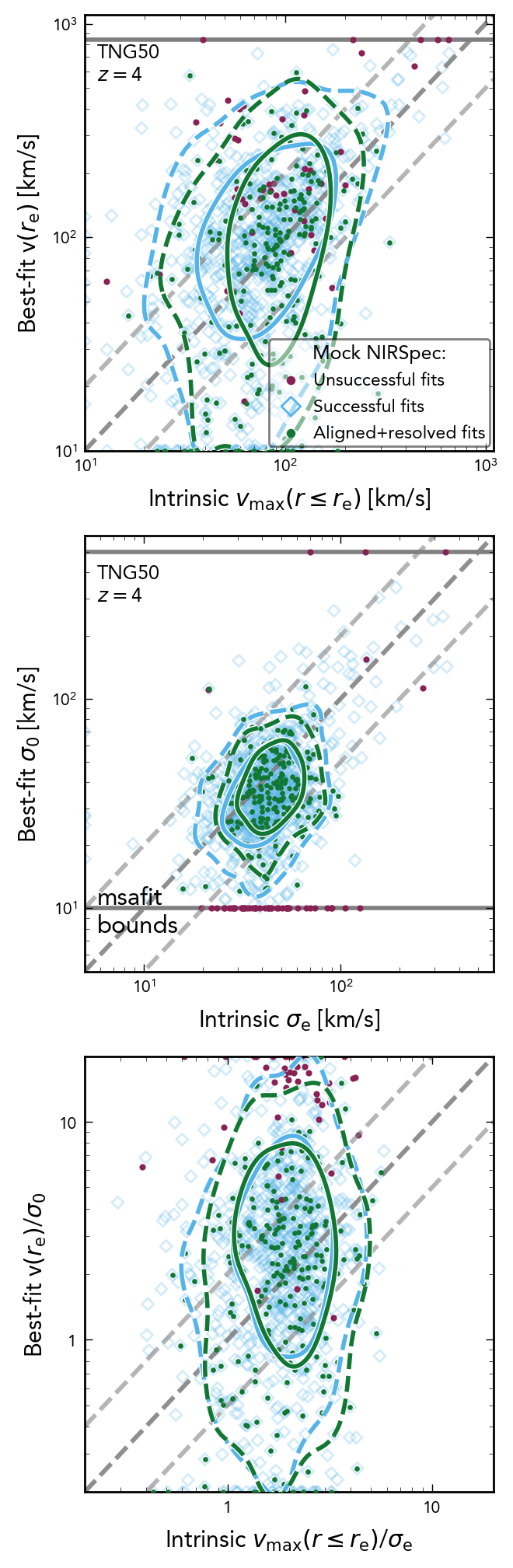}
    \includegraphics[width=0.32\textwidth]{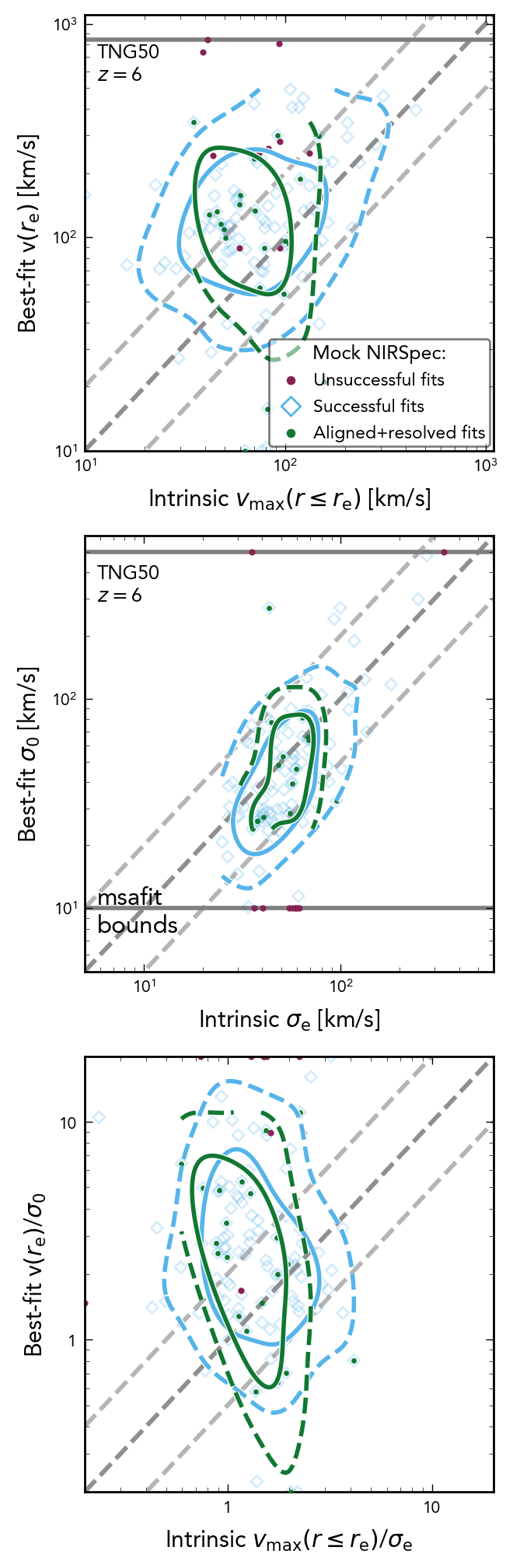}
    \caption{Differences in inference fidelity for $v$, $\sigma$ and $v/\sigma$: Best-fit \msavre\ vs intrinsic \TNGvre, best-fit \msasigma\ vs intrinsic \TNGsigma, and best-fit \msavre/\msasigma\ vs intrinsic \TNGvre/\TNGsigma\ for $z=2,4,6$. The lines, contours, and annotations are as in Fig.~\ref{fig:bestfitvsintrinsic}.} \label{fig:bestfitvsintrinsic_altz}
\end{figure*}

\begin{figure*}
    \includegraphics[width=0.42\textwidth]{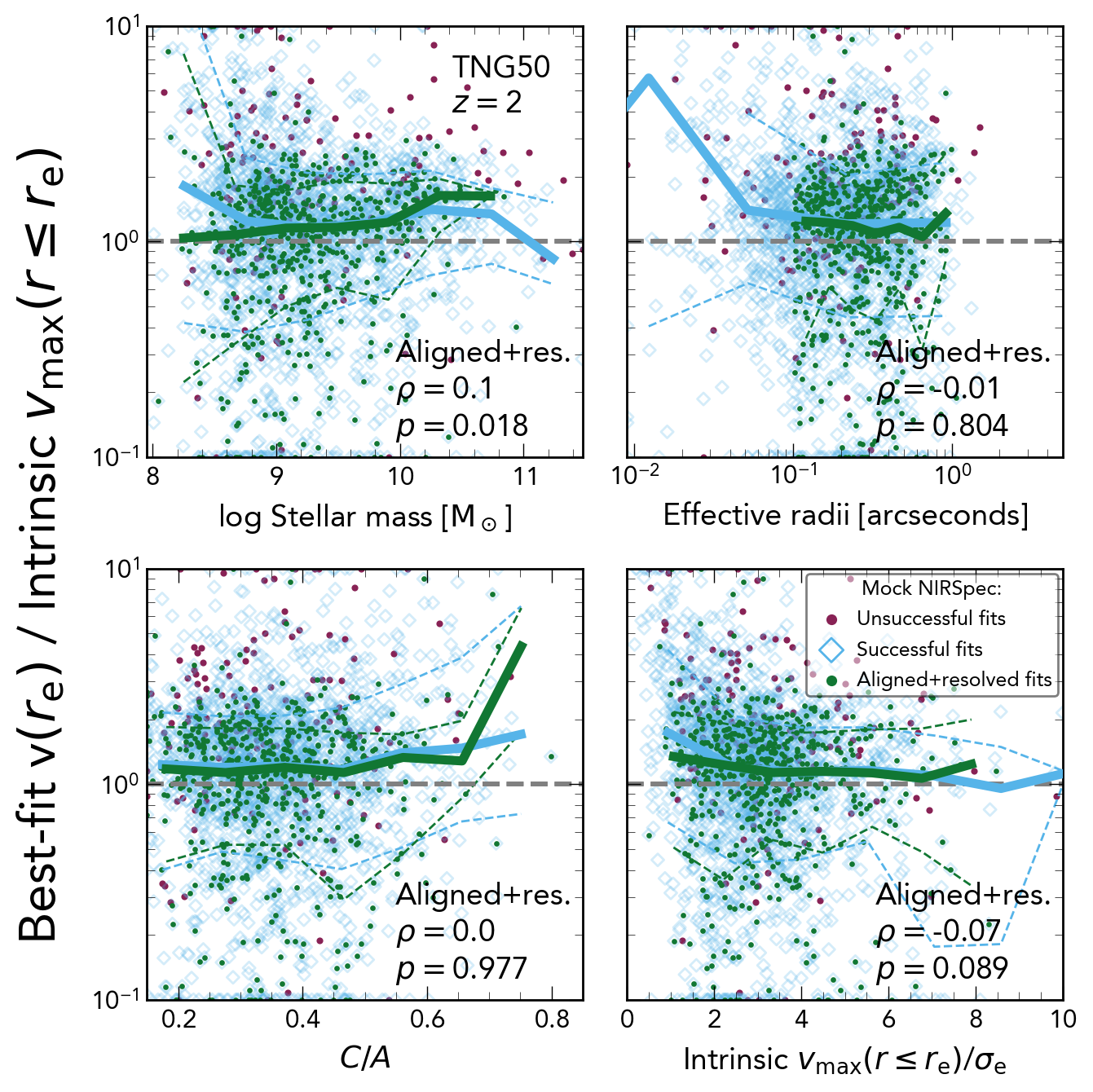}
    \includegraphics[width=0.42\textwidth]{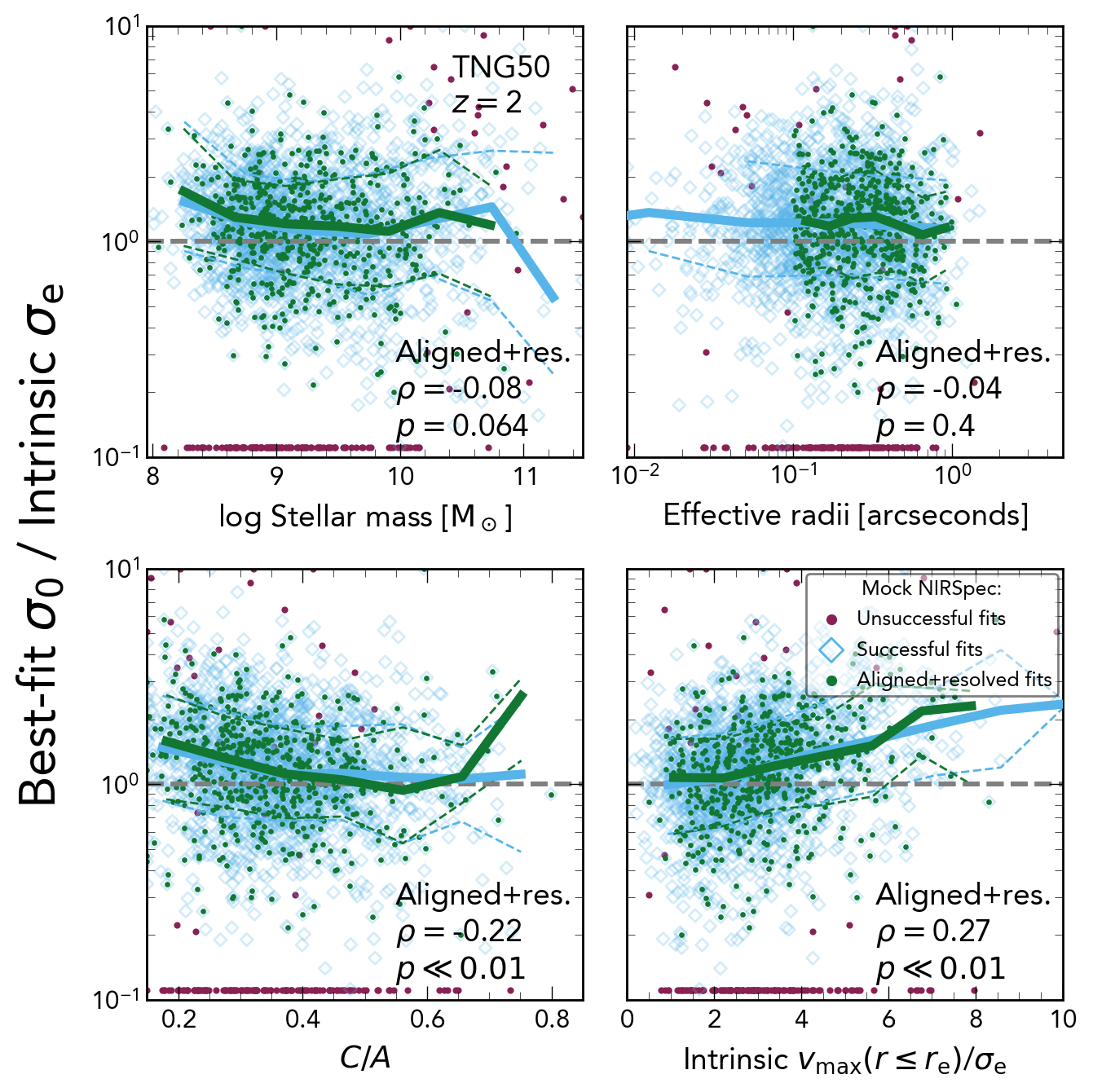}
    \includegraphics[width=0.42\textwidth]{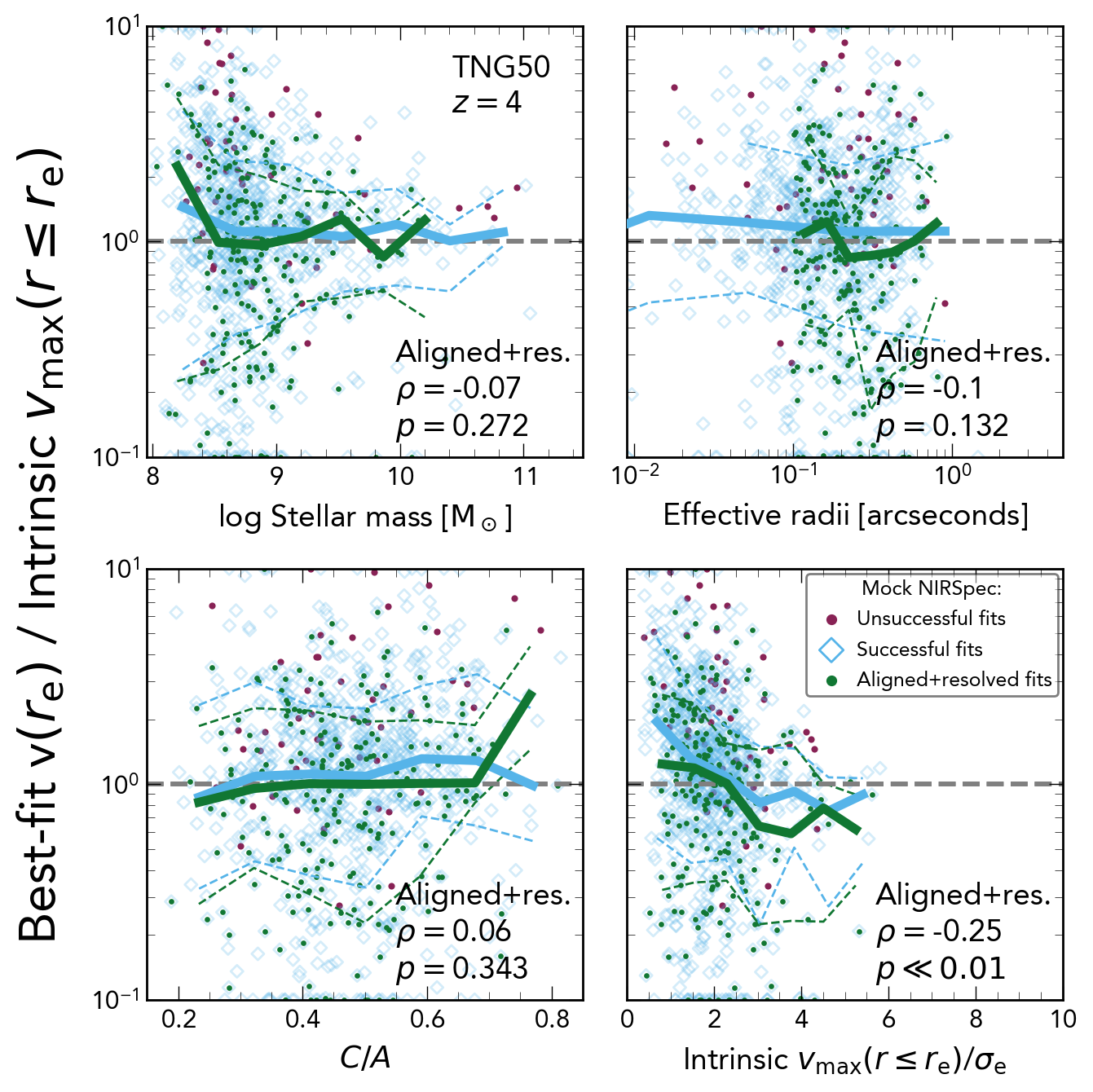}
    \includegraphics[width=0.42\textwidth]{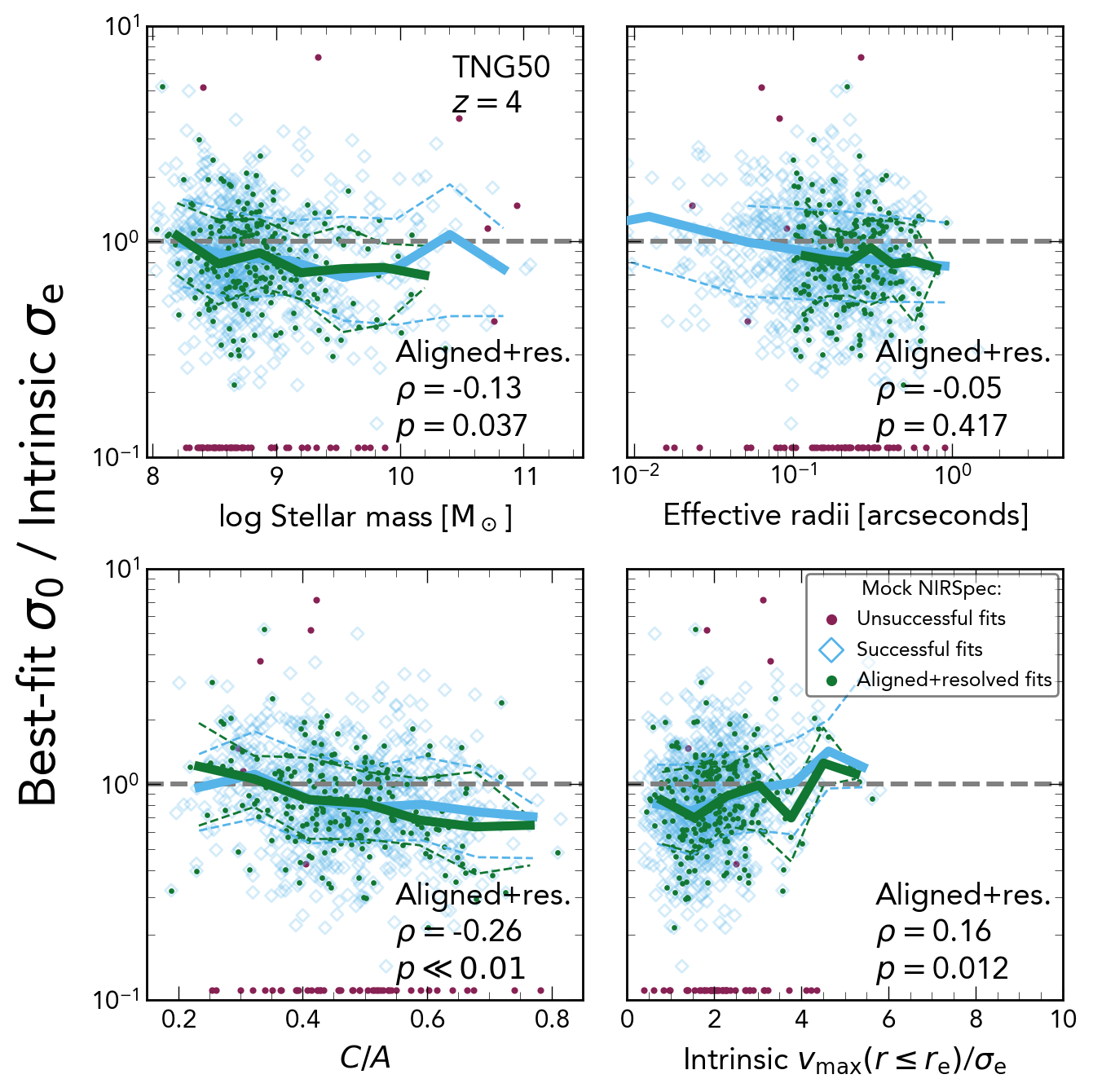}
    \includegraphics[width=0.42\textwidth]{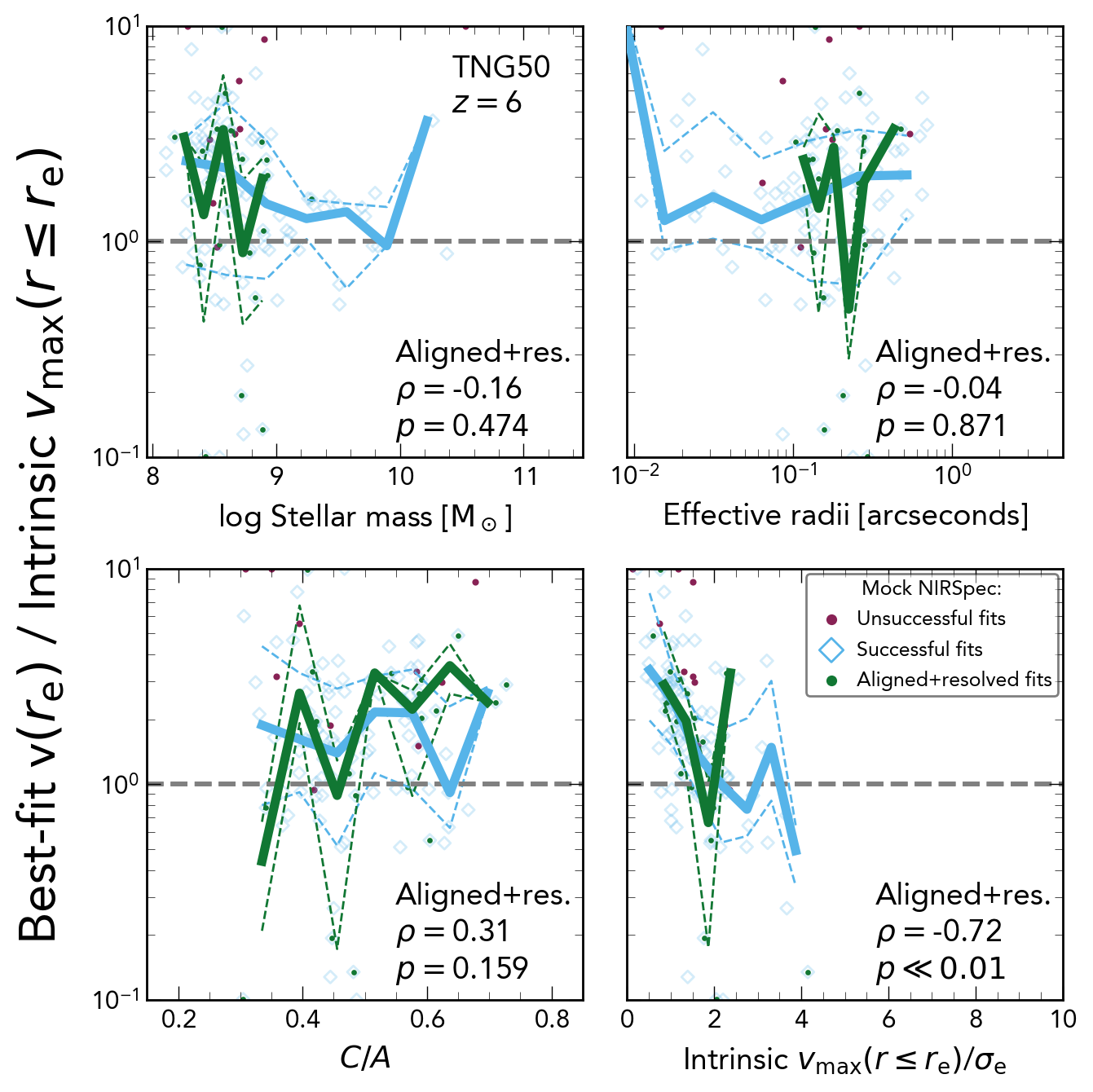}
    \includegraphics[width=0.42\textwidth]{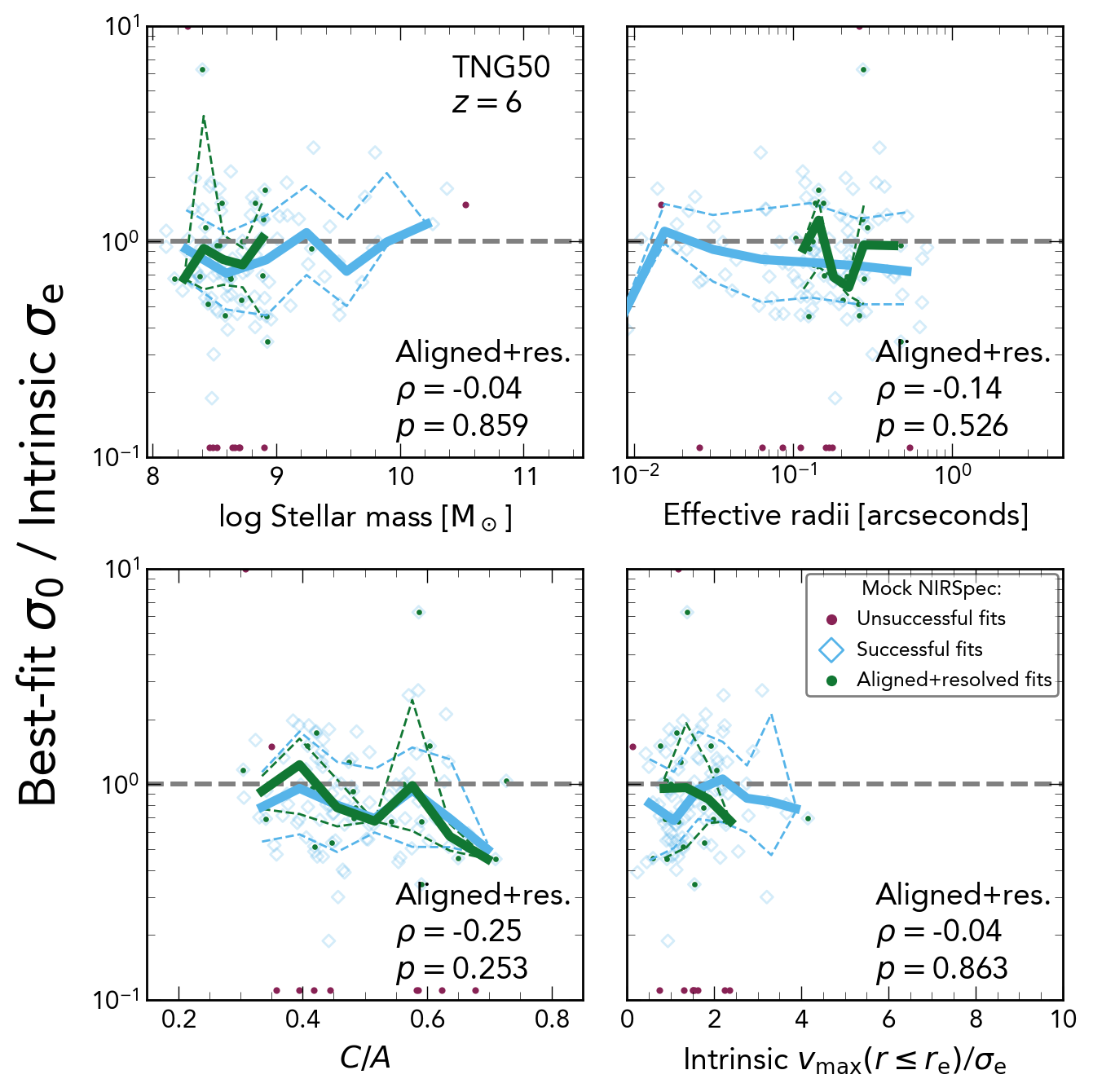}
    \caption{Differences in inference fidelity as a function of galaxy property: (Left) Best-fit \msavre\ by intrinsic TNG \TNGvre\ (Right) Best-fit \msasigma\ by intrinsic TNG \TNGsigma\ vs galaxy properties for $z=2,4,6$. The axes, lines, and annotations are as in Fig.~\ref{fig:bestfitbyintrinsic_vs_galprop}.}
    \label{fig:bestfitbyintrinsic_vs_galprop_altz}
\end{figure*}

In this Section, we compare best-fit to intrinsic kinematic properties $v$, $\sigma$ and $v/\sigma$ for TNG50 galaxies at $z=2,4,6$ in Fig.~\ref{fig:bestfitvsintrinsic_altz}. We also show the ratio of best-fit to intrinsic kinematics as a function of galaxy properties (stellar mass, effective radii, intrinsic shortest-to-longest axis ratios and \TNGvre/\TNGsigma) in Fig.~\ref{fig:bestfitbyintrinsic_vs_galprop_altz}.

\bsp	
\label{lastpage}
\end{document}